\documentclass[aps,pre,superscriptaddress,fleqn,showpacs,twocolumn]{revtex4}
\usepackage{amsmath,amssymb}
\usepackage{graphicx}
\usepackage{color}

\begin{document}

\title{Strong Effects of Network Architecture in the Entrainment of Coupled
Oscillator Systems}
\author{Hiroshi Kori}
\email[E-mail address: ]{kori@nsc.es.hokudai.ac.jp}
\affiliation{Department of Mathematics, Hokkaido university, Kita 10, Nishi 8, Kita-Ku,
Sapporo, Hokkaido, 060-0810, Japan}
\author{Alexander S. Mikhailov}
\affiliation{Abteilung Physikalische Chemie, Fritz-Haber-Institut der
Max-Planck-Gesellschaft, Faradayweg 4-6, 14195 Berlin, Germany}
\date{\today}

\begin{abstract}
Entrainment of randomly coupled oscillator networks by periodic external
forcing applied to a subset of elements is numerically and analytically
investigated. For a large class of interaction functions, we find that the
entrainment window with a tongue shape becomes exponentially narrow for
networks with higher hierarchical organization. However, the entrainment is
significantly facilitated if the networks are directionally biased, i.e.,
closer to the feedforward networks. Furthermore, we show that the networks
with high entrainment ability can be constructed by evolutionary
optimization processes. The neural network structure of the master clock of
the circadian rhythm in mammals is discussed from the viewpoint of our
results.
\end{abstract}

\maketitle

%05.45.Xt: Synchronization; coupled oscillators,
%89.75.Fb: Structures and organization in complex systems
%87.18.Sn: Neural networks
%biological clock

\section{introduction}

\label{sec:introduction} The study of complex networks has applications in
various fields including biology and engineering, and attracts growing
attention \cite{albert02,dorogovtsev03-1}. In the last decade, much progress
has been reached in understanding complexity of network architectures that
represent skeletons of networks. However, not only the architecture but also
the dynamics taking place in a network is important \cite{strogatz01}. This
dynamics determines \emph{functions} of the networks, related, e.g., to
information processing in the brain (which is a network of neurons) or to
the production process in a factory (which is a network of machines). Most
of the real-world networks are constructed in such a way that they implement
certain desired dynamical behaviors. Not only the individual components of a
network (such as single neurons), but also the network architecture (e.g.,
connections between neurons) should be appropriately designed. Thus, it
should be expected that real-world networks have architectures reflecting
their desired dynamical behaviors. If one can extract topological properties
of the networks relevant for their desired dynamics, this may provide
insights into the \emph{meaning} of network architecture from the viewpoint
of dynamics and help to understand the design principles of functional
networks.

In the present study, the focus is on the synchronization of oscillators
coupled into complex networks. Synchronization plays a crucial role in
functioning of various systems \cite
{winfree67,winfree80,kuramoto84,pikovsky01,manrubia04}. One of the most
intriguing examples is the circadian (i.e., approximately daily) clock
in mammals \cite{reppert02,aton-review05}. The circadian rhythms of the
whole body are orchestrated by a central clock in the brain, called the
suprachiasmatic nucleus (SCN). This brain tissue is formed by a
population of special neurons, known as clock cells. A single clock cell
exhibits a robust circadian rhythm in its firing rate and thus each such
cell is a self-sustained oscillator (with oscillations determined by a
regulatory loop in a group of genes at the single-cell level). In the
SCN, neurons mutually synchronize their physiological rhythms even in
absence of any environmental assistance \cite{yamaguchi03}, so that this
cell population can generate collective periodic signal. Additionally, a
different kind of synchronization takes place in the SCN which
represents entrainment to environmental rhythms. (Here and below, we use
the term \textquotedblleft entrainment\textquotedblright\ to describe
synchronization to an external input, distinguishing it from autonomous
mutual synchronization). The entrainment of the SCN is essential for the
proper functioning of this biological clock. Generally, the intrinsic
period of the circadian rhythm is significantly different from 24 hours
(in humans, it would typically be 25 h). Therefore, it must be tuned to
the 24 h period through some external influences. Moreover, the phase of
circadian oscillations needs to be locked appropriately to the local
time. The entrainment to the environment is mediated by the light
information coming from the eyes, which acts as external periodic
forcing for the SCN. However, only a distinct subset of neurons receives
and processes this light information, which implies that the rest of the
population should be indirectly entrained via intercellular
communication (see Ref.~\cite{reppert02} and references therein). There
are various types of intercellular communication inside the SCN,
including the communication through several types of neurotransmitters
\cite{aton-review05} . The communication through neurotransmitters goes
via chemical synapses of neurons forming a complex directed network.

With respect to mutual synchronization of oscillators, the impact of the
network architecture has been intensively investigated in recent years
\cite{barahona02,hong02,earl03,nishikawa03,ichinomiya04,ichinomiya05,restrepo05}.
However, only a few studies have dealt so far with the entrainment
(i.e., synchronization to external periodic forcing) of oscillator
networks \cite {yamada02,kori04,radicchi06}. The objective of the
present paper is to identify principal topological properties that
determine the entrainment ability of oscillator networks and to find the
dependence of the entrainment ability on these topological
properties. Although the problem is motivated by the SCN, our interest
is more general. Therefore, we use the phase oscillator model
\cite{kuramoto84}, which provides a good approximation of a broad class
of coupled oscillator systems and is expected to yield general insight
into the entrainment phenomena. In the earlier Letter, a particular
phase oscillator model has been studied and it was found that, in this
model, the entrainment threshold for the coupling intensity between
oscillators increases exponentially with the \emph{depth of a network},
defined as the mean forward distance from a pacemaker (i.e., source of
external forcing) to the network nodes \cite{kori04}. Hence, it was
found that the entrainment is practically only possible for the shallow
networks with low hierarchical organization. Here, we give all details
of the derivations, extend the analysis to different network models and
demonstrate the universality of the earlier preliminary
results. Additionally, a new class of random networks is considered,
which allows further to explore the design principles of entrainable
networks. The optimization of the network architecture with respect to
the entrainment through a dynamical learning process is further
performed.

The article is organized as follows: The model is introduced in Sec.~\ref%
{sec:model}. Then, we first show numerical results obtained for standard
random networks in Sec.~\ref{sec:numerical}. In Sec.~\ref{sec:analytical},
the model is analytically investigated under a certain approximation for the
network architecture. The comparison with numerical results is also
provided. In Sec.~\ref{sec:directivity}, a different class of random
networks is introduced, i.e. \emph{directionally biased networks}, and it is
shown that the entrainment is strongly enhanced for the networks close to
the feedforward-type. In Sec.~\ref{sec:evolution}, dynamical evolution of
the network architecture using two kinds of learning algorithms is
considered. The results are discussed, with a special emphasis on the design
principles of biological clocks, in Sec.~\ref{sec:discussion}, followed by
main conclusions in Sec.~\ref{sec:conclusion}.

\section{The models}

\label{sec:model} We consider a system of $N+1$ phase oscillators, one of
them being a pacemaker. The basic model is given by a set of evolution
equations for the oscillator phases $\phi _{i}$ and the pacemaker phase $%
\phi _{0}$, 
\begin{eqnarray}
\dot{\phi}_{i} &=&\omega +\frac{\kappa }{pN}\sum_{j=1}^{N}A_{ij}\Gamma (\phi
_{i}-\phi _{j})+\mu B_{i}\tilde{\Gamma}(\phi _{i}-\phi _{0}),  \notag \\
\dot{\phi}_{0} &=&\omega +\Delta \omega .  \label{model}
\end{eqnarray}%
The topology of network connections is determined by the adjacency matrix ${%
\mathbf{A}}$ whose elements $A_{ij}$ are either $1$ or $0$. The mean degree $%
pN$ is the average number of incoming connections per node (and $p$ is
called the connectivity). The element with $i=0$ is special and represents a
pacemaker. Its frequency is increased by $\Delta \omega $ with respect to
the frequency $\omega $ of all other oscillators. The pacemaker is acting on
the oscillators $1\leq i\leq N_{1}$, the action being specified by the
coefficients $B_{i}$ taking $1$ for $1\leq i\leq N_{1}$ and $0$ otherwise.
The coupling between elements inside the network is characterized by the
coupling function $\Gamma (x)$ and the (positive) coupling intensity
coefficient $\kappa $. In absence of a pacemaker, such networks undergo
autonomous phase synchronization at the natural frequency $\omega $ if the
coupling is attracting, i.e., if $\Gamma ^{\prime }(0)<0$. The coupling to
the pacemaker is characterized by the $2\pi $-periodic function $\tilde{%
\Gamma}(x)$ and the (positive) intensity coefficient $\mu $. We assume that
functions $\Gamma (x)$ and $\tilde{\Gamma}(x)$ are $2\pi $-periodic,
nonconstant, and smooth.

Without loss of generality, our model can be simplified. By going into a
rotating frame, we have $\omega =0$. The maxima of the coupling function, $%
\max \Gamma $ and $\max \tilde{\Gamma},$ are chosen equal to unity by
properly defining the coefficients $\kappa $ and $\mu $. Moreover, rescaled
time $t^{\prime }=t$ $\Delta \omega $ and rescaled coupling strengths $%
\kappa ^{\prime }=\kappa /\Delta \omega ,\mu ^{\prime }=\mu /\Delta \omega $
\ are introduced\footnote{%
We assume here $\Delta \omega >0$. However, the results in the present paper
hold also for $\Delta \omega <0$ because of the following reason. The system
(1) is invariant under transformation $\omega \rightarrow -\omega $, $\Delta
\omega \rightarrow -\Delta \omega $, $\phi \rightarrow -\phi $,$\Gamma (\phi
)\rightarrow -\Gamma (-\phi )$, and $\tilde{\Gamma}(\phi )\rightarrow -%
\tilde{\Gamma}(-\phi )$. It will be turned out that the entrainment behavior
does not depend on the explicit form of $\Gamma (\phi )$ provided that $%
\Gamma ^{\prime }(0)<0$. Therefore, the same entrainment behavior takes
place also for $\Delta \omega <0$. However, in this case, the role of $\max
\Gamma $ is replaced by $-\min \Gamma $, so that $-\min \Gamma $ should be
put unity by properly rescaling $\kappa $.}. After that, the model takes the
form of Eq.~(\ref{model}) with $\Delta \omega =1$ and $\omega =0$ (below, we
drop primes in the notations for the rescaled quantities). Note that, in
terms of the original model (\ref{model}), an increase of the rescaled
coupling between the elements corresponds either to an increase of coupling $%
\kappa $ or to a decrease of the relative pacemaker frequency $\Delta \omega 
$.

The presence of a pacemaker imposes hierarchical organization in the network
architecture, which plays a crucial role in determining the entrainment
ability. For any node $i$, its distance $h$ with respect to the pacemaker is
defined by the length of the minimum \emph{forward} path separating this
node from the pacemaker. Thus, the whole network is divided into a set of 
\emph{shells}, each of which is composed by oscillators with the distance $h$
from the pacemaker. The shell population $N_{h}$ is given by the number of
the oscillators with distance $h$. The \emph{depth} $L$ of a network is
introduced as 
\begin{equation}
L=\frac{1}{N}\sum_{h}hN_{h},
\end{equation}%
which is the average distance from the pacemaker to the entire network. We
may classify the types of connections into \emph{forward, backward}, and 
\emph{intra-shell connections}, which are, respectively, connections from
the nodes in a certain shell $h$ to the nodes in the next shell $h+1$, from
the nodes in a certain shell $h$ to the nodes in the shallower shells $k<h$,
and between the nodes inside the same shell. We call the connections coming
from a certain node the \emph{outgoing connections} of the node, and the
connections received by a certain node the \emph{incoming connections} of
the node.

In the analysis of the model, several further assumptions will be
made. The number $N_{1}$ of elements, directly connected to the
pacemaker, is assumed to be small as compared with the total size $N$ of
the network. The mean degree $pN$ is chosen large enough, so that the
networks do not become disconnected. We assume $\Gamma (0)=0$ which
implies that coupling between connected oscillators vanishes when their
phases are synchronized. Moreover, coupling of the elements to the
pacemaker is chosen to be much stronger than coupling between the
elements in the rest of the network ($\mu \gg \kappa $).  In all the
numerical simulations performed in the present paper, instead of using
Eq.~(\ref{model}) we put $\phi_i = \phi_0$ for $1\le i \le
N_1$, corresponding to the the limit $\mu \to \infty$.

\section{Numerical investigations for standard random networks}

\label{sec:numerical} We begin our analysis of entrainment phenomena by
considering the important special case of standard random networks, also
known as Erd\"{o}s-R\'{e}nyi (ER) networks
\cite{erdos59,bollobas85}. These networks are generated by independently
assigning with probability $p$ $\ $ for any pair $i$ and $j$ of the
network nodes a connection that leads from the node $i$ to the node
$j$. Hence, elements $A_{ij}$ of the adjacency matrix $\bf A$ are chosen
to be $1$ with probability $p$ and $0$ otherwise, and $\bf A$ is
asymmetric in general.  Only sparse random networks with a small mean
degree $pN$ $\ll N$ will be considered. We use the following coupling
function
\begin{equation}
\Gamma (x)=-\frac{\sin (x+\alpha )-\sin \alpha }{1+\sin \alpha },
\label{gamma-alpha}
\end{equation}%
where $\alpha $ is constant. Note that $\Gamma (0)=0$ and $\max \Gamma (x)=1$
for any $\alpha $. Note also that for $\alpha =0$ this coupling function is
the same as that used in Ref.~\cite{kori04}.

Numerical simulations are started with random phases for the oscillators $%
i>N_{1}$. For each oscillator, its effective long-time frequency $\omega
_{i} $ is computed as 
\begin{equation}
\omega _{i}=\frac{\phi _{i}(t_{0}+T)-\phi _{i}(t_{0})}{T},
\end{equation}%
with sufficiently large $T$ and $t_{0}$.

\begin{figure}[tbp]
\resizebox{8cm}{!}{\includegraphics{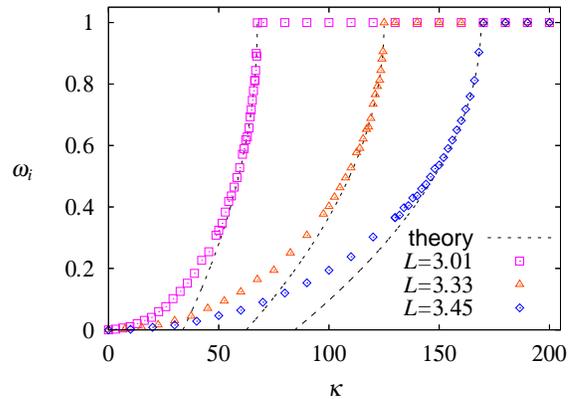}}
\caption{(color online). Long-time frequencies for various $\kappa $%
. Data obtained for three samples of random networks is plotted with
different symbols. The depths of the networks are $L=3.01,3.33$ and $3.45$.
The lines show the theoretical curves (\ref{omega-k}) with $%
\kappa _{\mathrm{cr}}=67.5,125$ and $169$ respectively. $N=100,p=0.1,N_{1}=1$%
, and $\alpha =0$.}
\label{fig:below}
\end{figure}

Numerical simulations show that the response of a network to the
introduction of a pacemaker depends on the strength $\kappa $ of coupling
between the oscillators. When this coupling is sufficiently large, the
pacemaker entrains the whole network (i.e., $\omega _{i}=1$ for any $i$).
Under entrainment, the relative phases $\phi _{i}-\phi _{0}$ are locked. As
the coupling strength $\kappa $ is decreased, the entrainment breaks down at
a certain threshold value $\kappa _{\mathrm{cr}}$ (for $\alpha =0$, see Fig.~%
\ref{fig:below}). Our simulations show that, in most cases, synchronization
between the first and the second shells is the first to break down, and the
frequencies of oscillators in the shells $h\geq 2$ remain equal for any $%
\kappa $, i.e. $\omega _{i}=\omega _{j}$ for $i,j>N_{1}$. The results for $%
\alpha =0.5$ and $\alpha =-0.5$ are similar. 

\begin{figure}[tbp]
\resizebox{8cm}{!}{\includegraphics{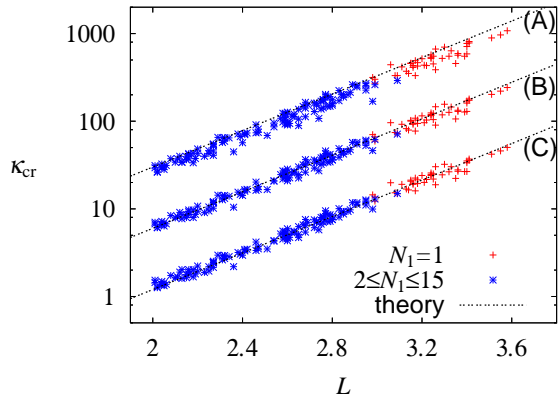}}
\caption{(color online). Dependences of the entrainment threshold on the
depth $L$ for a large set of random networks of size $N=100$ and $pN=10$.
The coupling function is given by Eq.~(\ref{gamma-alpha}) with (A) $%
\alpha =0.5$, (B) $\alpha =0$, and (C) $\alpha =-0.5$%
. To separate the data sets, here we have plotted $\nu %
\kappa _{\mathrm{cr}}$ with $\nu =5,1$ and $0.2$ respectively for
(A), (B) and (C). The lines are the theoretical dependence $c\nu 
\kappa _{\mathrm{cr}}$, where $\kappa _{\mathrm{cr}}$ is
given by Eq.~(\ref{kcr-fig}) and $c=0.60$ is a fitting parameter.}
\label{fig:kcr}
\end{figure}

Figure \ref{fig:kcr} displays the thresholds $\kappa _{\mathrm{cr}}$ (in the
logarithmic scale) for a large set of networks with the fixed mean degree $%
pN $ and different numbers $N_{1}$ of oscillators in the first shell. As can
be seen, the entrainment threshold $\kappa _{\mathrm{cr}}$ may vary largely
for different network realizations even if $N_{1}$ is fixed. The results for 
$\alpha =-0.5,0$, and $0.5$ are labeled as A, B and C. The same set of
networks is used for all three values of $\alpha $. For a given network, the
entrainment thresholds obtained for different values of $\alpha $ are close,
implying that the entrainment threshold does not depend on the particular
form of the coupling function. Each group of networks with a certain $N_{1}$
is displayed with its own symbol. Each group is characterized by a
distribution of depths $L$ and generates therefore a cluster of data points.
Correlation between the entrainment threshold $\kappa _{\mathrm{cr}}$ and
the network depth $L$ is evident even within a fixed $N_{1}$. The
distributions inside each cluster and the accumulation of the clusters yield
the dependence $\kappa _{\mathrm{cr}}(L)$ of the entrainment threshold on
the network depth. Remarkably, the observed dependence is well numerically
approximated by the exponential law 
\begin{equation}
\kappa _{\mathrm{cr}}\propto (1+pN)^{L}.  \label{exp-kcr}
\end{equation}

\begin{figure}[tbp]
\resizebox{8cm}{!}{\includegraphics{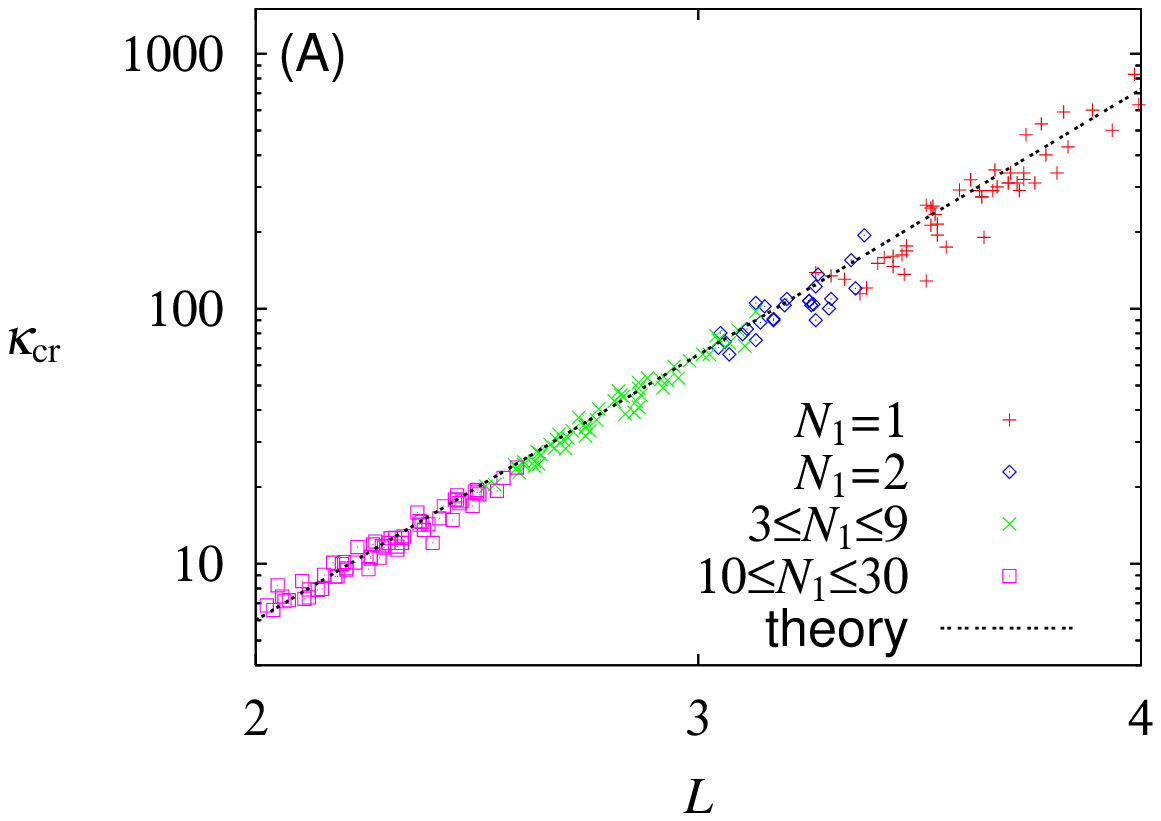}} 
\resizebox{8cm}{!}{
\includegraphics{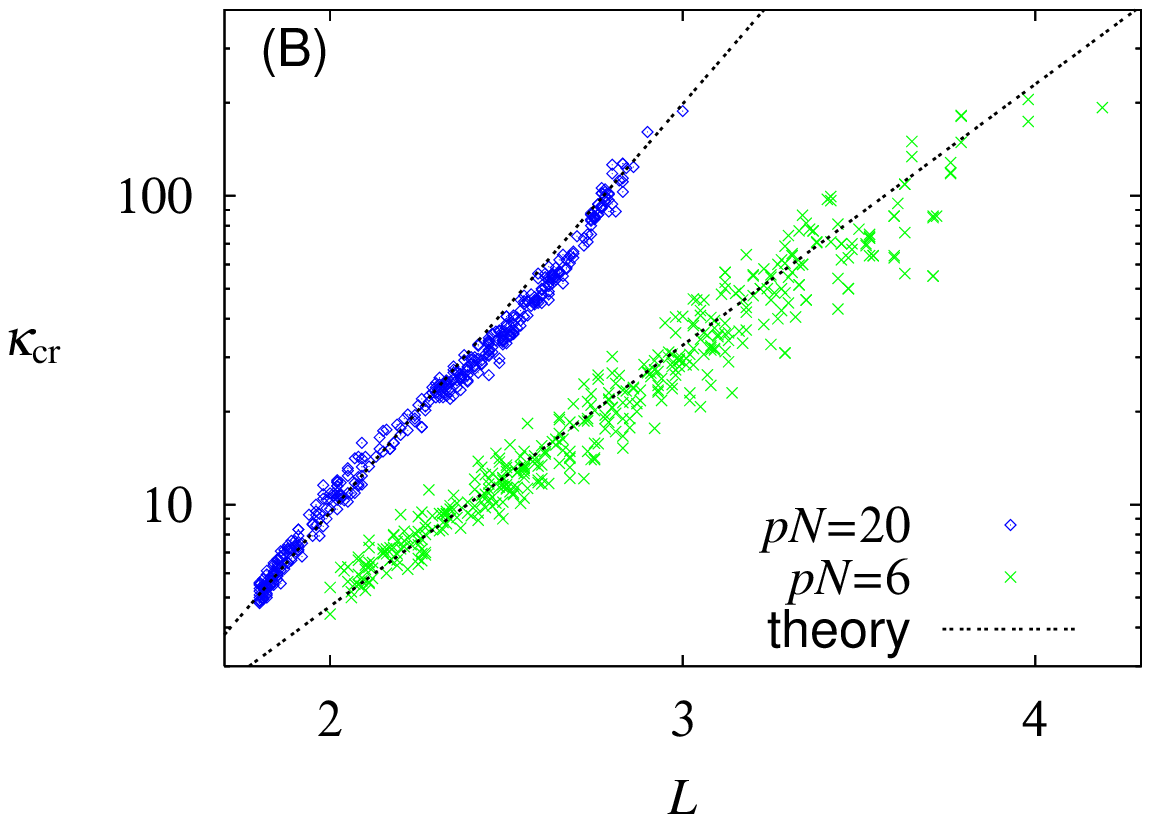}}
\caption{(color online). Dependences of the entrainment threshold on the
depth $L$ for a large set of random networks. The coupling function is given
by Eq.~(\ref{gamma-alpha}) where $\alpha =0$. (A) $N=200$
and $pN=10$ . The line is the the theoretical dependence $c\kappa _{%
\mathrm{cr}}$ , where $\kappa _{\mathrm{cr}}$ is Eq.~(\ref%
{kcr-fig}) and $c=0.60$ (the same as the value used for $N=100$). (B) $N=100$
and $pN=6,20$. The lines are the theoretical dependence $c\kappa _{%
\mathrm{cr}}$, where $\kappa _{\mathrm{cr}}$ is Eq.~(\ref%
{kcr-fig}) and $c=0.67$ and $0.45$ respectively for $pN=6$ and $20$.}
\label{fig:kcr-other}
\end{figure}

The same functional dependence is found for networks of other system sizes
and different average mean degrees $pN$ (see Fig.~\ref{fig:kcr-other}). The
entrainment thresholds for $N=100$ and $N=200$ can be well fitted by the
same function with the same fitting parameter $c$, which suggests that the
entrainment threshold depends not explicitly on the system size $N,$ but on
the average degree $pN$. 

\begin{figure}[tbp]
\resizebox{8cm}{!}{\includegraphics{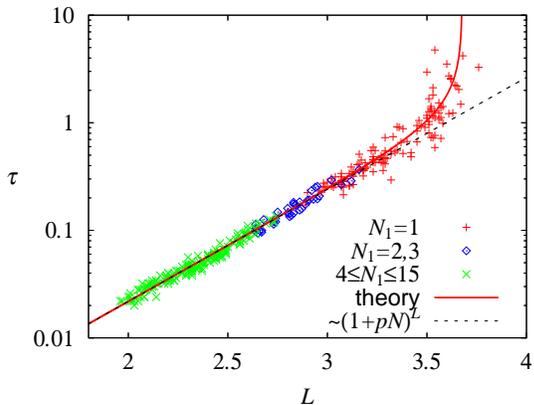}}
\caption{(color online). Dependence of the relaxation time on the depth $L$
for an ensemble of random networks. The parameters are $N=100$, $p=0.1$, $%
\alpha =0$, and $\kappa =300$. The solid line is the
theoretical dependence $d\tau $ where $\tau $ is Eq.~(%
\ref{tau2}) with $\kappa _{\mathrm{cr}}=0.6pN(1+pN)^{L-2}$
(same as the theoretical lines in Fig.~\ref{fig:kcr}), and $d=1.22$
is an additional fitting parameter. The dotted line is the exponential
function $\kappa _{\mathrm{cr}}/\kappa \propto (1+pN)^{L}$. }
\label{fig:tau}
\end{figure}

Next we consider relaxation to the entrained state. The relaxation time for
each generated network has been measured as follows. First, we have run a
numerical simulation for a long time starting with a random initial
distribution of phases. If the relative phases $\psi _{i}\equiv \phi
_{i}-\phi _{0}$ for all $i$ were not locked at the end of the simulation
(implying that the network did not become entrained for the chosen coupling
intensity), such a network has been discarded. Then, the simulation was
repeated starting from the initial condition $\phi _{i}(t=0)=\psi _{i}^{%
\mathrm{st}}+\epsilon y_{i}$ for $i>N_{1}$, where $y_{i}$ is a random number
independently taken from the uniform distribution within $(0,1),$ $\epsilon $
is a small coefficient and $\psi _{i}^{\mathrm{st}}$ are the relative phases
of oscillators in the entrained state. The time dependence of the distance $%
D(t)\equiv \sum_{i}\sqrt{\{\psi _{i}(t)-\psi _{i}^{\mathrm{st}}\}^{2}}$ has
been monitored. The relaxation time $\tau $ was defined as the time $t$ at
which the ratio $D(t)/D(0)$ becomes equal to $e^{-1}$. Figure~\ref{fig:tau}
displays the relaxation times $\tau $ (in the logarithmic scale), obtained
numerically for a large set of networks and for a fixed coupling strength $%
\kappa $. Again, exponential dependence on the depth is evident. This
dependence is well fitted by the function 
\begin{equation}
\tau _{\mathrm{cr}}\propto (1+pN)^{L}.  \label{exp-tau}
\end{equation}%
There is a divergence of the relaxation time around $L=3.7$. This divergence
occurs around the region of the depth where given coupling strength $\kappa $
is close to the entrainment threshold, i.e., $\kappa \simeq \kappa _{\mathrm{%
cr}}(L)$. Above this region, entrainment rarely happens. 

\begin{figure}[tbp]
\resizebox{8cm}{!}{\includegraphics{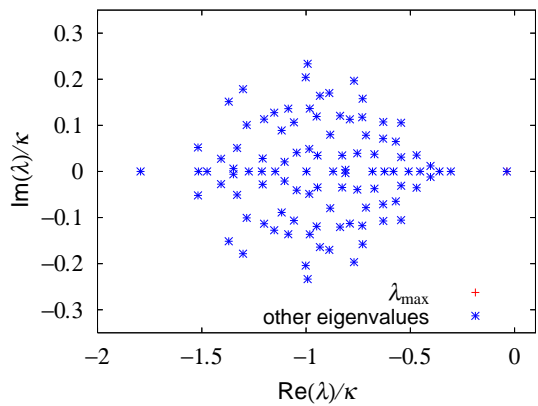}}
\caption{(color online). Eigenvalues (divided by $\kappa $) of the
entrained state. The parameters are $N=100$, $N_{1}=2$, $p=0.1$, $%
\alpha =0$, and $\kappa =300$ (the depth of the used network is $%
L=2.86$ and the entrainment threshold for this network is found numerically
to be $\kappa _{\mathrm{cr}}\simeq 33$). The real part of the
maximum eigenvalue is $\mbox{\rm Re}(\lambda _{\mathrm{max}}) -11$.}
\label{fig:eigen}
\end{figure}

Eigenvalues of an entrained state are displayed in Fig.~{\ref{fig:eigen}}.
They were obtained by numerically solving the stability matrix of our model (%
\ref{model}). We have assumed the limit $\mu \rightarrow \infty $ and
perturbations are considered only in the subsystem $i>N_{1}$, so that there
are $N-N_{1}$ eigenvalues. The eigenvalue possessing the maximum real part
is denoted by $\lambda _{\mathrm{max}}$. As seen, the magnitude of the real
part of this eigenvalue, $|\mbox{\rm Re}\lambda _{\mathrm{max}}|$, is much
smaller than those of others (which are distributed around $|\kappa |$). We
have preliminarily checked the eigenvectors and found that the associated
eigenvector with $\lambda _{\mathrm{max}}$ corresponds to an approximately
identical phase shift of the whole subsystem $i>N_{1}$, and the rest
eigenvectors correspond to relative motions inside this subsystem. This
suggests the following relaxation process: Relative perturbations inside the
subsystem quickly diminish within the characteristic time scale $\kappa ^{-1}
$ and, then, the phase differences between the oscillators in the subsystem
become practically locked. The relaxation of the phase difference between
the first shell and the subsystem proceeds slowly with the time scale $|%
\mbox{\rm Re}\lambda _{\mathrm{max}}|^{-1}$. Such behavior has actually
been observed in our numerical simulations. 

The results relating to the relaxation process do not change qualitatively
for different values of $\alpha $ (we have further checked them for $\alpha
=0.5$ and $\alpha =-0.5$). Note also that the relaxation time does not
depend on the particular form of small perturbations.

\section{Analytical investigations in the global tree approximation}

\label{sec:analytical} In this Section, the model is analytically
investigated using the heuristic global tree approximation. First, global
tree networks will be introduced and then their similarity to the ER
networks will be explained. After that, the analytical solution of the
entrainment problem for global tree networks will be constructed. Note that
this analytical solution will hold only in the limit when both $pN$ and $N$
are large.

\subsection{Global tree approximation}

\begin{figure}[tbp]
\resizebox{8cm}{!}{\includegraphics{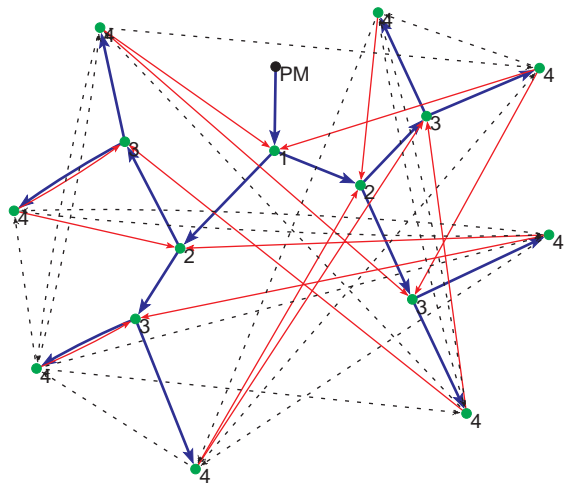}}
\caption{(color online). An example of a global tree network ($N_1=1, pN=2,
H=4$, and hence, $N=15$). Thick solid lines, thin solid lines, and dotted
lines represents respectively forward connections, backward connections, and
intra-shell connections. Numbers indicate the hierarchical level of each
node. This figure is drawn by Pajek (http://vlado.fmf.uni-lj.si/pub/networks/pajek/).}
\label{fig:tree}
\end{figure}

A global tree network is a hierarchical network where, in any level, each
oscillator has only one incoming connection from the higher level and
exactly $pN$ outgoing connections leading to the lower level. Additionally,
each oscillator has exactly $pN$ connections from the last (bottom) shell of
this hierarchical network. No other connections exist and the precise
connection topology remains arbitrary. An example of a global tree network
with $4$ hierarchical levels is shown in Fig.~\ref{fig:tree}. 

For such networks, shell populations grow exponentially, i.e. as $%
N_{h}=N_{1}(pN)^{h-1}$ with the distance $h$ from the network origin. The
total number $H$ of shells is determined by the condition $%
\sum_{h=1}^{H}N_{h}=N$. The depth of such a network is given by $%
L=N^{-1}\sum_{h=1}^{H}hN_{h}=H+O(1/pN)$. Thus, for $pN$ $\rightarrow \infty $%
, the depth $L$ coincides with the number of shells $H$. Therefore, $H$ can
be approximately replaced by $L$.

Global tree networks share essential properties with the ER networks of
large size $N\gg 1$ and high connectivity $pN\gg 1.$ This similarity is
briefly explained below (see Appendix~\ref{sec:withouttree} for a detailed
discussion). We first consider the pattern of forward connections of the ER
network. By definition of our model, each node in the first shell receives
one forward connection from the pacemaker. Each node typically gives $pN$
outward connections. Thus, from the first shell, a total number of the
outward connections is $pNN_{1},$ on the average. All elements receiving a
connection from the first shell form the second shell of the network. If the
number of connections leading from the first shell is much smaller than the
total number $N$ of elements to which they may lead (i.e., if $pNN_{1}\ll N$%
), each next outward connection from the first shell is typically received
by a different node in the second shell. This means that, typically, an
element in the second shell would be linked only to a single element in the
first shell, as required by the tree structure. Considering the third shell,
we can notice that again that, if the number $N_{3}=N_{1}(pN)^{2}$ of
outgoing connections from this shell is small as compared with the network
size $N$, the tree structure would approximately hold for this shell too.

Shell populations $N_{h}$ grow exponentially with the number $h$ of the
shell, i.e. $N_{h}=N_{1}(pN)^{h-1}$. The tree structure with respect to
forward connections holds so far as these populations remain much
smaller than the total network size. When $pN$ is large, it can be shown
that only the last two shells have populations of size $O(N)$ and, thus,
the tree structure approximately holds down to the third last
shell. Analyzing further patterns of backward and intra-shell
connections in a large ER network, we notice that populations of all
shells, except the last of them, are of order $o(N)$ and only the
populations of the two last shells are of order $O(N)$. Therefore, each
node outside of the last two shells receives backward connections mostly
from these last two shells. On the average, the number of such
connections is $pN$. Each node in the last two shells typically has $pN$
incoming connections from the last two shells. This large number of
connections between the last two shells strongly facilitates
synchronization of oscillators inside them, as can be seen in numerical
simulations. Having this in mind, we may merge the last two shells in
the network into a single shell. Once this additional (empirical)
approximation is used, the tree structure is extended to the entire
network, including its last shells.

\begin{figure}[tbp]
\resizebox{8cm}{!}{\includegraphics{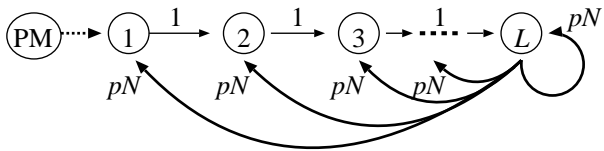}} \caption{Schematic
representation of the network structure along a forward path in the global
tree approximation. Each circle with a number $h$ is an oscillator of the
shell $h$. Each right-arrow and each left-arrow represent respectively a
forward connection and $pN$ backward (or, intra-shell) connections. }
\label{fig:1d-tree}
\end{figure}

Thus, in the \emph{global tree approximation} the ER network is treated
as a network having a tree structure with respect to forward connections
and with the backward connections arriving only from the last network
shell.  In this network, every forward path starting from the pacemaker
has a similar structure, as illustrated in Fig.~\ref{fig:1d-tree}.

When this approximation is used, analytical solution can be easily
constructed and its properties become more clear. Therefore, we prefer first
to present our analysis using the global approximation, even also it remains
partly empirical. The investigation for ER networks can also be performed
without relying on the tree approximation. While giving quantitatively
better results, this other analytical investigation is however more
complicated and, therefore, we give it separately in Appendix A.

\subsection{The entrained solution}

\label{sec:entrain} In the considered tree networks, our model (\ref{model})
can be written as 
\begin{equation*}
\dot{\phi}_{i}^{1}=\mu \tilde{\Gamma}(\phi _{i}^{1}-t)+\frac{\kappa }{pN}%
\sum_{j}A_{ij}\Gamma (\phi _{i}^{1}-\phi _{j}^{L}),
\end{equation*}%
\begin{equation}
\dot{\phi}_{i}^{h}=\frac{\kappa }{pN}\Gamma (\phi _{i}^{h}-\phi _{k}^{h-1})+%
\frac{\kappa }{pN}\sum_{j}A_{ij}\Gamma (\phi _{i}^{h}-\phi _{j}^{L}),
\label{model2}
\end{equation}%
where $h\geq 2$; $\phi _{i}^{h}$ is the phase of the oscillator $i$ in
the shell $h$; $\phi _{k}^{h-1}$ is the oscillator giving forward
connection to the oscillator $i$ in the shell $h$; and the summation is
taken over all the oscillator $j$ in the last shell $L$. Because each
oscillator in any particular shell has the same pattern of connections,
i.e., one forward connection from the previous shell and $pN$ backward
connections from the last shell. Thus, the entrained state of a network
with phase synchronization inside every shell is possible.  For such a
state, the entrainment solution problem is reduced into that in the
one-dimensional oscillator array along a forward path starting from the
pacemaker (see Fig.~\ref{fig:1d-tree}).  Putting $\dot{\phi}_{i}=1$
(which is the entrainment condition) and $\phi _{i}^{h}=\theta _{h}$ for
all the oscillator $i$ in the shell $h$, we get the following algebraic
equations:
\begin{equation}
\mu \tilde{\Gamma}(\theta _{1}-t)+\kappa \Gamma (\theta _{1}-\theta _{L})=1,
\label{reducedmodel1}
\end{equation}%
\begin{equation}
\frac{\kappa }{pN}\Gamma (\theta _{h}-\theta _{h-1})+\kappa \Gamma (\theta
_{h}-\theta _{L})=1\quad \mbox{for $h \ge 2$}.  \label{reducedmodel2}
\end{equation}%
For large $pN$, we may linearize $\Gamma (\theta _{h}-\theta _{k})$ as $%
\Gamma ^{\prime }(0)(\theta _{h}-\theta _{k})$ for $h,k\geq 2$ [it will be
shown that $\theta _{2}-\theta _{L}$ is at most of order $O(1/pN)$ in the
solution under entrainment, so that this linearizion is justified]. We may
then solve Eq.~(\ref{reducedmodel2}) from the last shell upward. For $h=L$,
we get 
\begin{equation}
\Gamma (\theta _{L}-\theta _{L-1})=\kappa ^{-1}pN.
\end{equation}%
Substitution this into Eq.~(\ref{reducedmodel2}) for $h=L-1$ results in 
\begin{equation}
\Gamma (\theta _{L-1}-\theta _{L-2})=\kappa ^{-1}pN(1+pN).
\end{equation}%
From the following shells, we need to employ the liniearlization, i.e., we
approximate 
\begin{equation}
\Gamma (\theta _{h}-\theta _{L}) = -\sum_{k=h-1}^{L}\Gamma (\theta
_{k+1}-\theta _{k})\quad \mbox{for $h\ge 2.$}
\end{equation}%
We then obtain 
\begin{equation}
\Gamma (\theta _{h}-\theta _{h-1})=\kappa ^{-1}pN(1+pN)^{L-h}\quad 
\mbox{for
$h \ge 2$}.  \label{phase}
\end{equation}%
Substituting Eq.~(\ref{phase}) for $h=2$ into Eq.~(\ref{reducedmodel1}), we
also obtain 
\begin{equation}
\tilde{\Gamma}(\theta _{1}-t)=\frac{1-\kappa \Gamma (\theta _{1}-\theta _{L})%
}{\mu }\leq \frac{1-\kappa ~\min \{\Gamma (x)\}}{\mu }.  \label{upper-bound}
\end{equation}%
Note that the explicit expression for $\tilde{\Gamma}(\theta _{1}-t)$ is not
needed.

The existence conditions of the entrained solution are $\tilde{\Gamma}%
(\theta _{1}-t)\leq 1$ and $\Gamma (\theta _{h}-\theta _{h-1})\leq 1$ for $%
h>2$. The former condition is satisfied for $\mu \gg \kappa $ (which is our
assumption). Among the terms $\Gamma (\theta _{h}-\theta _{h-1})$, the term $%
\Gamma (\theta _{2}-\theta _{1})$ is the largest one. The solution thus
exists if 
\begin{equation}
\Gamma (\theta _{2}-\theta _{1})\leq 1.
\end{equation}%
Therefore, the entrained state of the network is possible only if the
coupling intensity $\kappa $ satisfies condition $\kappa \geq \kappa _{%
\mathrm{cr}}$, where 
\begin{equation}
\kappa _{\mathrm{cr}}=pN(1+pN)^{L-2}.  \label{kcr}
\end{equation}

The result (\ref{kcr}) is remarkable. According to it, the coupling
threshold $\kappa _{\mathrm{cr}}$ is determined exclusively by topological
properties and does not depend on a particular form of the coupling
function. Moreover, it is given only through a combination of $pN$ and $L$.
This fact suggests that the essential parameters of the system are $pN$ and $%
L$ rather than $N,p$ and $N_{1}$, in agreement with the previous numerical
results obtained for the ER networks.

From Eq.~(\ref{kcr}), $\kappa _{\mathrm{cr}}$ for our tree network under
consideration is roughly estimated as 
\begin{equation}
\kappa _{\mathrm{cr}}\sim (pN)^{L-1} \sim N/N_{1}.  \label{kcr-estimate}
\end{equation}%
Thus, $\kappa _{\mathrm{cr}}$ is a large number for small $N_{1}$ (as we
assumed). This property will be used in the derivation of the relaxation
time.

One can verify that $\theta _{2}-\theta _{L}$ is at most of order $(pN)^{-1}$
in the entrained state by substituting $\kappa _{\mathrm{cr}}$ into $\kappa $
in Eq.~(\ref{phase}). Note also that although the phase differences $\theta
_{h}-\theta _{h+1}$ are small in deeper shells, they should not be neglected
when the entrainment threshold is derived, because the phase difference $%
\theta _{1}-\theta _{2}$ (which determines the existence condition of the
entrained solution) is a consequence of the exponential growth of such small
phase differences from the deepest shell upwards.

\subsection{Stability analysis of the entrained state}

\label{sec:stability} To analyze stability of the entrained solution,
only the limit $\mu \rightarrow \infty $ is considered (so that $\phi
_{i}=\theta _{1}\rightarrow t$ for $i\leq N_{1}$). In numerical
simulations, we have observed that the relaxation process is
characterized by two distinct time scales: the first one characterizes
fast relaxation inside the subsystem $i>N_{1}$ and the other corresponds
to slow relaxation in the phase difference between the first shell and
the subsystem. In this section, we first investigate the \emph{internal}
stability of the subsystem. Then, taking advantage of the separation of
time scales, we heuristically construct the effective dynamical equation
of the whole subsystem.

We consider small perturbations for phases in the entrained state, i.e., $%
\phi _{i}^{h}=\theta _{h}+\delta \phi _{i}^{h}$ where $\delta \phi _{i}^{h}$
is small perturbation. Because of the assumption $\mu \rightarrow \infty $,
we put $\delta \phi _{i}^{1}=0$, i.e., $\phi _{i}^{1}=\theta _{1}=t$. The
model (\ref{model2}) for $h\geq 2$ then gives 
\begin{eqnarray}
\delta \dot{\phi }_{i}^{h} &=&\frac{\kappa }{pN}\Gamma ^{\prime }(\theta
_{h}-\theta _{h-1})(\delta \phi _{i}^{h}-\delta \phi _{k}^{h-1})  \notag
\label{model3} \\
&&+\frac{\kappa }{pN}\sum_{j}A_{ij}\Gamma ^{\prime }(\theta _{h}-\theta
_{L})(\delta \phi _{i}^{h}-\delta \phi _{j}^{L}).
\end{eqnarray}%
Because $\theta _{h}-\theta _{L}=O(1/pN)$, we approximate $\Gamma ^{\prime
}(\theta _{h}-\theta _{L})\simeq \Gamma ^{\prime }(0)$. In addition, because
of the large number $pN$ of connections from the last shell, we may
approximate 
\begin{equation}
\frac{1}{pN}\sum_{j}A_{ij}\delta \phi _{j}^{L}\simeq \frac{1}{N_{L}}%
\sum_{j}\delta \phi _{j}^{L}\equiv \overline{\delta \phi ^{L}}.
\end{equation}%
Using these two approximations, Eq.~(\ref{model3}) reduces to 
\begin{equation}
\delta \dot{\phi }_{i}^{h}=\frac{\kappa }{pN}\Gamma ^{\prime }(\theta
_{h}-\theta _{h-1})(\delta \phi _{i}^{h}-\delta \phi _{k}^{h-1})+\kappa
\Gamma ^{\prime }(0)(\delta \phi _{i}^{h}-\overline{\delta \phi ^{L}}).
\label{model4}
\end{equation}%
Since the first term is negligibly small for large $pN$ for any shell $h\geq
2$, Eq.~(\ref{model4}) is further approximated as 
\begin{equation}
\delta \dot{\phi }_{i}^{h}\simeq \kappa \Gamma ^{\prime }(0)(\delta \phi
_{i}^{h}-\overline{\delta \phi ^{L}}).  \label{model5}
\end{equation}%
Thus, in our approximation, all eigenvalues associated with the relative
motion inside the subsystem are degenerated into 
\begin{equation}
\lambda =\kappa \Gamma ^{\prime }(0),  \label{lambda}
\end{equation}%
with degeneracy $N-N_{1}-1$. The dynamical equation (\ref{model5}) describes
the fast relaxation inside the subsystem: oscillators quickly relaxes to the
average perturbation $\overline{\delta \phi ^{L}}$ in the last shell, i.e., $%
\phi _{i}^{h}\rightarrow \theta _{h}+\overline{\delta \phi ^{L}}$. After
this fast relaxation, the phase differences inside the subsystem are almost
locked.

Now we heuristically construct the dynamical equation for the phase
difference between the first shell and the subsystem. We take the coordinate
of the subsystem on its \emph{surface}, i.e., the phase of the second shell,
defined as $\Psi = \theta_2 + \overline{\delta\phi^L}$. Because the total
external force applied to the subsystem is given by $N_2\kappa
\Gamma(\Psi-t)/pN = N_1 \kappa \Gamma(\Psi-t)$, the effective dynamical
equation for the surface $\Psi$ reads 
\begin{equation}  \label{subsystem}
N_{\mathrm{e}} \dot \Psi = N_1\kappa \Gamma(\Psi-t).
\end{equation}
Here, the \emph{effective size} $N_{\mathrm{e}}$ of the whole network is
determined by the condition that $\Psi=\theta_2$ when the subsystem is
entrained (i.e. $\dot \Psi=1$). From this condition, comparing Eq.~(\ref%
{phase}) for $h=2$ and Eq.~(\ref{subsystem}), we obtain 
\begin{equation}  \label{effective-size}
N_{\mathrm{e}} = N_1 \kappa_{\mathrm{cr}}.
\end{equation}
Thus, from Eqs.~(\ref{subsystem}) and (\ref{effective-size}) the last
eigenvalue is found to be 
\begin{equation}  \label{lambda-max}
\lambda _{\mathrm{max}} =\frac{\kappa \Gamma ^{\prime }(\theta _{2}-\theta
_{1})}{\kappa _{\mathrm{cr}}}.
\end{equation}
The eigenvector of $\lambda_{\mathrm{max}}$ corresponds to the identical
phase shift for all the oscillators $i>N_{1}$.

For the phase difference $\theta _{2}-\theta _{1}$, there are at least one
pair of steady solutions, one of which always satisfies $\Gamma ^{\prime
}(\theta _{2}-\theta _{1})<0$. Thus, provided that $\Gamma ^{\prime }(0)<0$,
a stable entrained solution exists for $\kappa<\kappa_{\mathrm{cr}}$. This
implies that the entrainment breakdown occurs only via the disappearance of
the solution at $\kappa =\kappa _{ \mathrm{cr}}$.

Because $\kappa_{\mathrm{cr}}$ is a large number for small $N_1$ [see Eq.~(%
\ref{kcr-estimate})], the time scales are well separated, i.e., $|\lambda_{%
\mathrm{max}}| \ll |\lambda|$. This fact justifies the approximation
employed in this subsection. In addition, the relaxation time against any
general perturbation is thus given simply by 
\begin{equation}
\tau = |\lambda_{\mathrm{max}}|^{-1} = - \frac{ \kappa_{\mathrm{cr}}}{\kappa
\Gamma^{\prime}(\theta_2-\theta_1)}.  \label{tau}
\end{equation}
Hence, in general, the relaxation time depends on the explicit form of the
coupling function, being different from the case of the entrainment
threshold. Nevertheless, the dependence on the $L$ is approximately the same
as $\kappa_{\mathrm{cr}}$ in the region of $L$ where $\Gamma^{\prime}(%
\theta_2-\theta_1)$ does not vary much with $L$.

For the special case $\Gamma (x)=-\sin x$, Eq.~(\ref{lambda-max}) reduces
into 
\begin{equation}
\lambda _{\mathrm{max}}=-\frac{\sqrt{\kappa ^{2}-\kappa _{\mathrm{cr}}^{2}}}{%
\kappa _{\mathrm{cr}}}.  \label{lambda-max2}
\end{equation}%
The relaxation time is thus 
\begin{equation}
\tau =\frac{\kappa _{\mathrm{cr}}}{\sqrt{\kappa ^{2}-\kappa _{\mathrm{cr}%
}^{2}}}.  \label{tau2}
\end{equation}%
Because $\kappa _{\mathrm{cr}}$ increases with $L$, $\kappa _{cr}$
eventually coincides with $\kappa $ at certain critical $L$, which results
in the divergence of $\tau $. In other region of $L$, the dependence of the
relaxation time on the depth $L$ is approximately the same as $\kappa _{%
\mathrm{cr}}$, i.e. exponential.

\subsection{Below the entrainment threshold}

\label{sec:below} Dynamical behavior just below the entrainment threshold is
considered. We again choose the limit $\mu \rightarrow \infty $. Because of
the property $|\lambda _{\mathrm{max}}|\ll |\lambda |$, we only need to
consider the dynamics of the subsystem, Eq.~(\ref{subsystem}). We introduce
the slow mode $x\equiv \Psi -t$ and the bifurcation parameter $\epsilon
\equiv (\kappa -\kappa _{\mathrm{cr}})/\kappa _{\mathrm{cr}}$. Substituting
them into Eq.~(\ref{subsystem}), we obtain 
\begin{equation}
\dot{x}=(1+\epsilon )\Gamma (x)-1.
\end{equation}%
Expansion of $\Gamma (x)$ up to the second order around its maximum $x=x_{%
\mathrm{max}}$ yields 
\begin{equation}
\dot{x}=\epsilon +\frac{\Gamma _{\mathrm{max}}^{\prime \prime }}{2}(x-x_{%
\mathrm{max}})^{2}+\mbox{higher orders},  \label{dotx}
\end{equation}%
where $\Gamma _{\mathrm{max}}^{\prime \prime }<0$ is the second derivative
of $\Gamma (x)$ at $x=x_{\mathrm{max}}$. The steady solution of $x$
(corresponding to the entrainment) thus disappears via a saddle-node
bifurcation at $\epsilon =0$, and $x$ oscillates with the negative frequency 
$\omega _{x}$ for $\epsilon <0$. This frequency is found through the
integration of Eq. (\ref{dotx}): 
\begin{equation}
\frac{2\pi }{\omega _{x}}=\int_{0}^{2\pi }\frac{dx}{\epsilon +\Gamma _{%
\mathrm{max}}^{\prime \prime }(x-x_{\mathrm{max}})^{2}/2}+O(\epsilon ),
\label{integral}
\end{equation}%
followed by $\omega _{x}=-\sqrt{2\epsilon /\Gamma _{\mathrm{max}}^{\prime
\prime }}$. Hence, the effective frequency $\omega _{i}$ of each oscillator
in the subsystem ($i>N_{1}$) for $\epsilon <0$ is 
\begin{equation}
\omega _{i}=1-\sqrt{\frac{2\epsilon }{\Gamma _{\mathrm{max}}^{\prime \prime }%
}}+O(\epsilon ).  \label{omega-k}
\end{equation}

\subsection{Comparison with numerical results}

The analytical results obtained under the global tree approximation are
compared with the numerical results obtained for the ER networks. The
theoretical curves plotted in Fig.~\ref{fig:below} correspond to the
function (\ref{omega-k}) for appropriate values of $\kappa _{\mathrm{cr}}$,
showing excellent agreement with the numerical data for $\kappa $ close to $%
\kappa _{\mathrm{cr}}$ , as expected. In Figs.~\ref{fig:kcr} and \ref%
{fig:kcr-other}, we display 
\begin{equation}
\kappa _{\mathrm{cr}}=cpN(1+pN)^{L-2},  \label{kcr-fig}
\end{equation}%
where $c$ is a fitting parameter added to our theoretical dependence (\ref%
{kcr}). As already noticed in Ref.~\cite{kori04}, although the principal
dependence on the depth, $(1+pN)^{L}$, is correctly reproduced by the global
tree approximation, the coefficient is somewhat different. We have thus
introduced the fitting parameter $c$. As seen in Figs.~\ref{fig:kcr} and \ref%
{fig:kcr-other}, the principal dependence on the depth obtained for the tree
network agrees excellently with the numerical results obtained for the ER
networks. Note that the theoretical dependence, analytically derived
directly for the ER network in Appendix~\ref{sec:withouttree}, agrees well
with numerical data without any fitting parameter. For the relaxation time,
taking into account the correction for $\kappa _{\mathrm{cr}}$, we use
instead of Eq.~(\ref{tau2}) the following function 
\begin{equation}
\tau =d\frac{c\kappa _{\mathrm{cr}}}{\sqrt{\kappa ^{2}-(c\kappa _{\mathrm{cr}%
})^{2}}},  \label{tau3}
\end{equation}%
where the value of $c$ is obtained by numerical fitting for the
entrainment threshold and $d$ is an additional fitting parameter. As
seen in Fig.~\ref {fig:tau}, the function (\ref{tau3}) fits very well to
numerical data. In Fig.~\ref{fig:eigen}, real parts of the eigenvalues
except $\lambda _{ \mathrm{max}}$ are scattered around $\kappa $, as
expected from Eq.~(\ref {lambda}) [note that $\Gamma ^{\prime }(0)=1$
for $\alpha =0$]. Putting $ \kappa =300$ and $\kappa _{\mathrm{cr}}=33$
into Eq.~(\ref{lambda-max2}), we obtain $\lambda_{\rm max} \simeq -9.0$,
which is close to the numerical result shown in Fig.~\ref{fig:eigen}
($\lambda _{\mathrm{max}}\simeq -11$).

\section{Entrainment in directionally biased networks } \label{sec:directivity}

We can understand by looking at Eq.~(\ref{reducedmodel2}) why entrainment
is difficult in the networks with larger depths. The first term on the left
side in the equation describes the force of the forward connection. Its
sign is positive, so that it contributes to the entrainment. On the other
hand, the second term describes the force from the backward connections,
whose sign is \textrm{negative}. By moving this second term to the right
side, it is seen that it essentially increases the frequency to which the
oscillators synchronize. In other words, the backward connections act as a
load for the entrainment. Moreover, the number ($pN$) of the backward
connections is much larger than the number (one) of the forward connections.
Thus, to compensate such strong unbalance, the phase difference associated
with the forward connection ($\theta _{h-1}-\theta _{h}$) needs to be much
larger that that associated with the backward connections ($\theta
_{h}-\theta _{L}$). The effect accumulates exponentially along a forward
path of the length $L$ starting from the pacemaker and ending at the last
shell. This accumulation results in the exponential growth the phase
difference from the last shell upwards, and thus, the exponential dependence
of the entrainment threshold on the depth $L$. Hence, due to the large
number of backward connections, the entrainment is very difficult for the
networks with large depths.

It is thus expected that the entrainment threshold decreases significantly
for the networks that are closer to the feedforward architecture (i.e., to a
network without backward connections). This is indeed demonstrated below
using a special class of random networks which we call \emph{directionally
biased networks}. To construct a directionally biased network, we first
generate a directed ER random network and choose $N_{1}$ nodes as the first
shell. Then, we redefine its backward connections. Namely, for every
backward connection, we decide to retain it with probability $\xi $ or
delete otherwise. We call $\xi $ the \emph{backward connectivity} of the
network. Note that a directed ER network corresponds to $\xi =1$ and the
feedforward network is obtained for $\xi =0$.

To solve the entrainment problems for such networks, the same approximations
as in Sec.~\ref{sec:analytical} are applied. The global tree approximation
for the forward connection pattern can be used. Furthermore, it can be
assumed that all backward connections come from the last shell and that the
number of such connections received by a oscillator is $\xi pN$. The former
approximation is applicable when $pN$ is large and the latter is applicable
when $\xi pN$ is large. Thus, we require that $\xi pN$ is large. Under this
condition, the same linear approximation as in Sec.~\ref{sec:analytical} can
be used, because the phase difference $\theta _{2}-\theta _{L}$ turns out to
be at most of order $(\xi pN)^{-1}$. The entrained solution for the
approximated network is found by solving algebraic equations 
\begin{equation}
\mu \tilde{\Gamma}(\theta _{1}-t)+\kappa \xi \Gamma (\theta _{1}-\theta
_{L})=1,  \label{reducedmodel3}
\end{equation}%
\begin{equation}
\frac{\kappa }{pN}\Gamma (\theta _{h}-\theta _{h-1})+\kappa \xi \Gamma
(\theta _{h}-\theta _{L})=1\quad \mbox{for $h \ge 2$}.  \label{reducedmodel4}
\end{equation}%
We obtain 
\begin{equation}
\Gamma (\theta _{h}-\theta _{h-1})=\kappa ^{-1}pN(1+\xi pN)^{L-h}\quad %
\mbox{for $h\ge 2$},  \label{phase-xi}
\end{equation}%
\begin{equation}
\kappa _{\mathrm{cr}}=pN(1+\xi pN)^{L-2}.  \label{kcr-xi}
\end{equation}%
The stability analysis is performed in the same manner, leading to 
\begin{equation}
\tau =\frac{\kappa _{\mathrm{cr}}}{\kappa \Gamma ^{\prime }(\theta
_{2}-\theta _{1})}.  \label{tau-xi}
\end{equation}%
Hence, both the entrainment threshold $\kappa _{\mathrm{cr}}$ and the
relaxation time $\tau $ decrease dramatically as the backward connectivity $%
\xi $ gets smaller. The role played by $\xi $ is more significant for the
networks with a higher hierarchical organization (i.e., with a larger depth $%
L$).

The employed linear approximation is applicable only for large $\xi
pN$. For $\xi =0$, the model is however exactly solvable without the
linear approximation. Since there are no backward connections in this
case, perfect phase synchronization inside each shell is possible.
Consequently, we get Eqs.~(\ref{phase-xi}) and (\ref{kcr-xi}) also for
$\xi =0$. Trivially, the phase difference $\theta _{h-1}-\theta
_{h}\equiv \Delta $ between the neighboring shell is constant for $h>2$,
given by the relation $\Gamma (-\Delta )=pN/\kappa$. The stability
analysis of this solution is straightforward. The eigenvalues are $\mu
\tilde{\Gamma}^{\prime }(\theta _{1}-t)$ (multiplicity $N_{1}$) and
$\kappa \Gamma ^{\prime }(-\Delta )/pN$ (multiplicity $N-N_{1}$). For
$\kappa >\kappa _{\mathrm{cr}}$, there is at least one pair of
solutions, one of which satisfies $\Gamma ^{\prime }(-\Delta
)<0$. Therefore, for any $\Gamma $, there is always a stable solution
for $\kappa >\kappa _{\mathrm{cr}}$. Note that this state has a constant
phase gradient and can thus be described as a traveling wave.

\begin{figure}[tbp]
\resizebox{8cm}{!}{\includegraphics{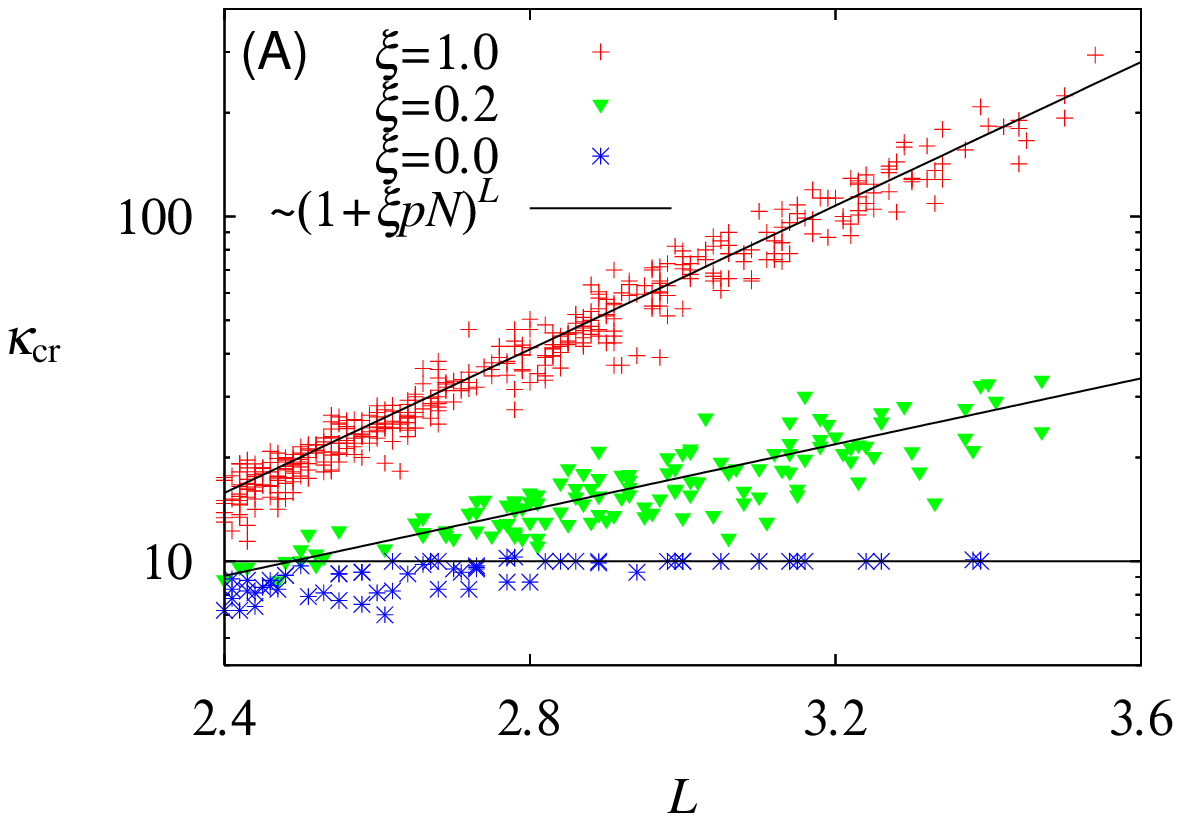}} \resizebox{8cm}{!}{%
\includegraphics{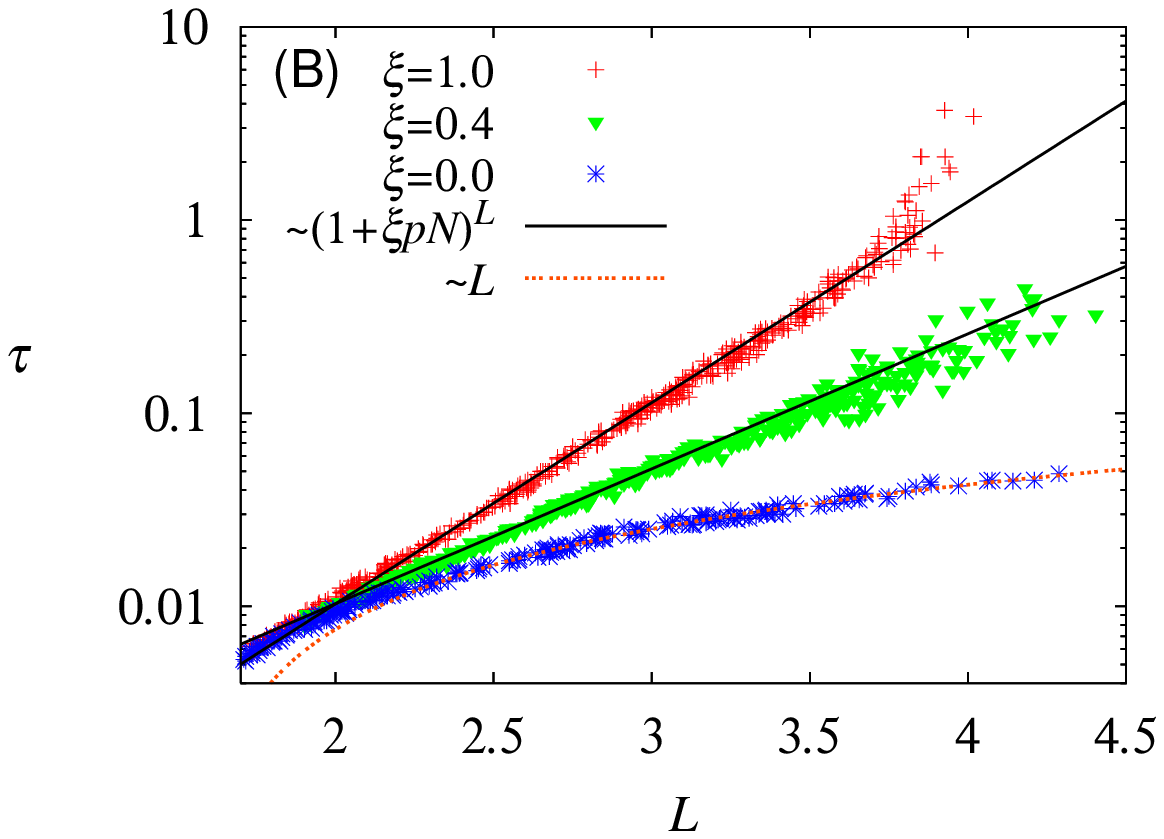}}
\caption{(color online). Dependences of (A) the entrainment threshold and
(B) the relaxation time on the depth for an ensemble of directionally biased
networks. The solid lines are the exponential functions proportional to $(1+%
\xi pN)^{L}$. The dash line is a linear function $aL+b$ with
appropriate fitting parameters $a$ and $b$. Other parameters are $%
\kappa =600,pN=10$, (A)$N=100$ and (B)$N=400$. . }
\label{fig:kcr-xi}
\end{figure}

Numerical results for directionally biased random networks are displayed in
Fig.~\ref{fig:kcr-xi}. For the entrainment thresholds (Fig.~\ref{fig:kcr-xi}%
A), the theoretical dependence $(1+\xi pN)^{L}$ fits well the numerical
data. For the relaxation time (Fig.~\ref{fig:kcr-xi}B), the theoretical
dependence $(1+\xi pN)^{L}$ also fits well the numerical data, except for $%
\xi =0$ where a linear function fits best. This linear dependence is natural
because the solution is wave-like and it is thus expected that the time
necessarily to transmit information is proportional to the system length
(i.e., to the depth $L$). Note that the entrainment breakdown occurs near $%
L=4$ for $\xi =1$. 

Similar dependences can be obtained for the \emph{weighted} networks. To
construct them, we first generate an ER network and introduce connections
from the pacemaker. We then put $A_{ij}=\xi $ for all existing backward
connections, while keeping the weights of other connections equal to either $%
0$ or $1$. For such networks, we also obtain Eqs.~(\ref{phase-xi})-(\ref%
{tau-xi}) because the same approximations are applicable. 

\section{Evolutionary optimization}

\label{sec:evolution} So far, the relationship between the entrainment
threshold \ and the network depth has been established only for the special
kinds of random networks. Therefore, it is natural to ask whether it also
holds for the arbitrary complex networks. Obviously, we cannot answer it by
investigate all possible types of complex networks because their number is
too large. Instead, an alternative way is chosen: we shall construct
networks with lower entrainment thresholds (and thus better entrainment
ability) through an optimization process starting from an arbitrary network
architecture and analyze changes of their topological properties in the
course of evolution. If these networks actually evolve towards being
shallower and closer to a feedforward type, one can conclude that both
properties are generally essential for good entrainment ability.

During the evolution, several topological properties shall be
monitored. One of them is the depth $L$, which has already been
defined. The other property is the backward connectivity $\xi $ defined
as the ratio of the total number of backward connections at each trial
and that of the initial network. For comparison, we also introduce the
\emph{forward connectivity} $\chi $, given by the ratio between the
total number of forward connections in each trial and that of the
initial network. The total number of outgoing connections from the first
shell, denoted by $n_{\mathrm{out}}$, shall also be monitored.

Two types of optimization algorithms will be employed. The first one is
the straightforward optimization where the network structure evolves in
such a way that the mean frequency of the \emph{whole} network becomes
closer to that of the pacemaker. In the second algorithm, each
oscillator selects its incoming connections in such a way that its
\emph{own} frequency becomes higher.  It will be shown that, for both
kinds, the evolving network improves its ability to become entrained
and, at the same time, indeed becomes less hierarchical and closer to
feedforward networks. Note that, concerning biological clocks, the first
and the second algorithms could imitate respectively the evolution
process of the neural network of the SCN for a species and the growing
process of the neural network for an individual. A more biological
algorithm (or, learning process) has been employed elsewhere and
resulted in similar results \cite{masuda06}.

The initial setup is common for both types. The process starts from a random
ER network. We put $\phi _{i}(t)=t$ for $i\leq N_{1}$. Throughout the
evolution process, the connection pattern from the pacemaker is maintained,
so that the first shell is always composed of the oscillators $i\leq N_{1}$.
The coupling strength $\kappa $ is fixed far below the entrainment threshold
for the initial network. Thus, all the oscillators except in the first shell
initially have frequencies close to their natural ones.

\emph{The first evolution algorithm}. A numerical simulation is run starting
from random initial phases. The long-time frequency $\langle \omega \rangle
=(TN)^{-1}\sum_{i}\left[ \phi _{i}(2T)-\phi _{i}(T)\right] ,$ averaged over
the whole population and for sufficiently long time $T$, is determined.
Then, the adjacency matrix $\mathbf{A}$ is mutated. This is done by
\textquotedblleft rewiring\textquotedblright : We choose randomly an
existing directed link ($A_{ij}=1$) and eliminate it ($A_{ij}\rightarrow 0$%
), and then, choose randomly a missing link ($A_{i^{\prime }j^{\prime }}=0$
where $i^{\prime }\neq j^{\prime }$) and add a new connection there ($%
A_{i^{\prime }j^{\prime }}\rightarrow 1$). After the mutation, a numerical
simulation is again run starting from new random initial phases and the new
frequency $\langle \omega \rangle ^{\prime }$ is determined. If $\langle
\omega \rangle ^{\prime }$ is closer to $1$ than $\langle \omega \rangle $,
the mutation is accepted. Otherwise, the network is reset to its structure
before the mutation. The iteration process is repeated until the long-time
frequency $\langle \omega \rangle $ becomes equal to unity.

\begin{figure}[tbp]
\resizebox{8cm}{!}{\includegraphics{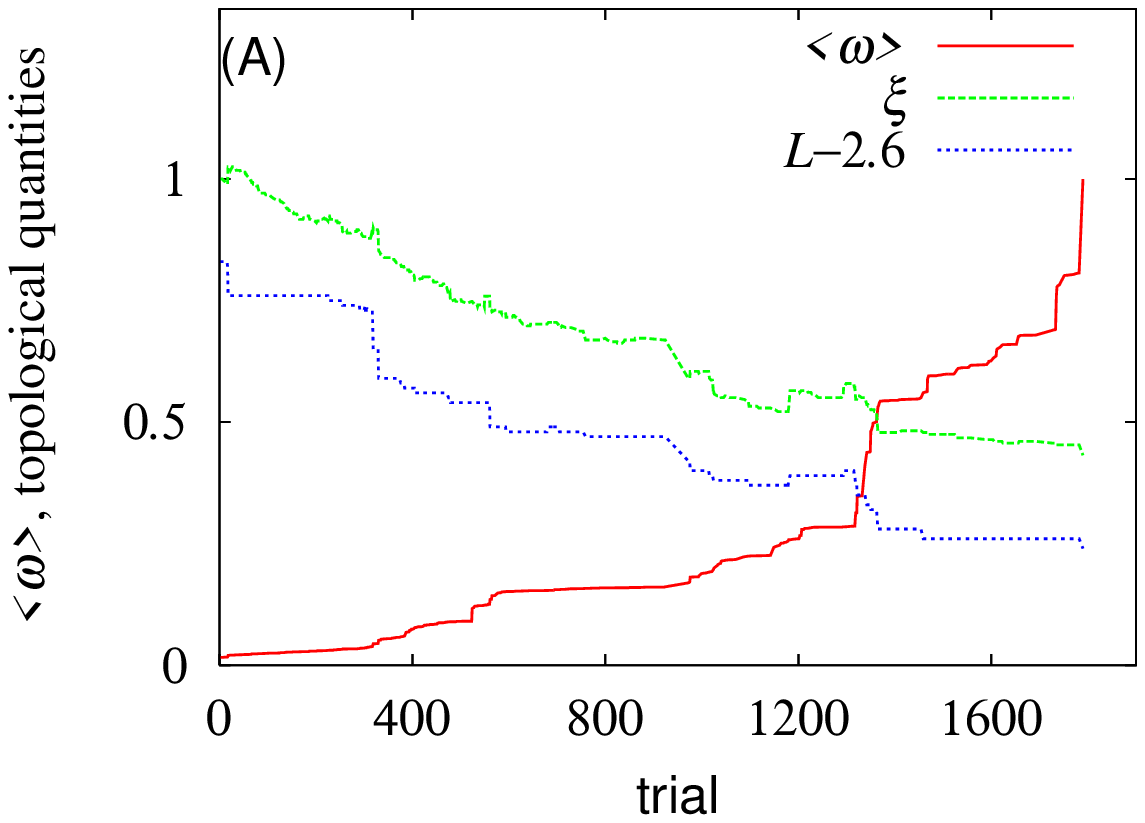}} \resizebox{8cm}{!}{%
\includegraphics{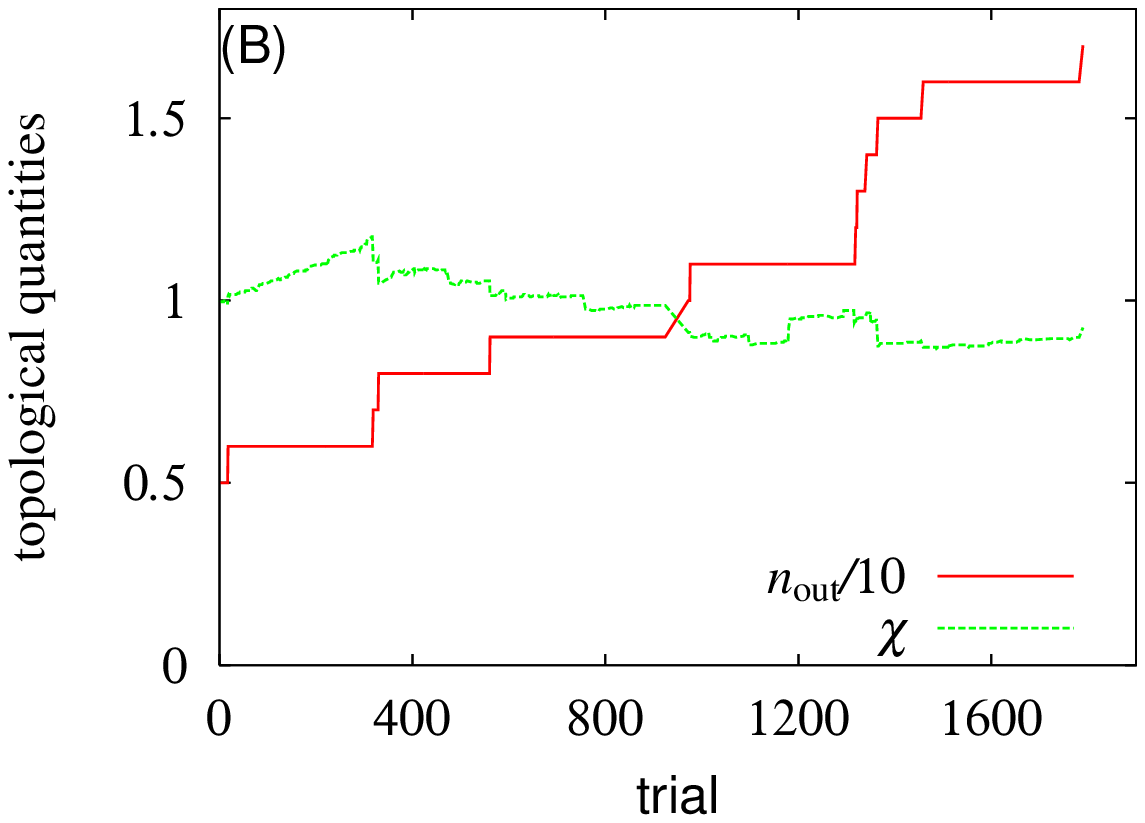}}
\caption{(color online). Evolution process under the first algorithm.
Parameter values are $N=100,p=0.1,N_{1}=1$ and $\kappa =20$. The
depth of the initial random network is $L=3.43$. (A)The solid line is the
long-time frequency $\langle \omega \rangle $ averaged over the
whole network. The dash line and the dot line are respectively the backward
connectivity $\xi $ and the depth $L$. (B)Two other topological
quantities are plotted. The solid line and the dot line are respectively the
total number $n_{\mathrm{out}}$ of outgoing connections from the first shell
and the forward connectivity $\chi $.}
\label{fig:learn}
\end{figure}

A typical dependence of the average frequency and two different topological
properties during an evolution is shown in Fig.~\ref{fig:learn}A. The
tendency to decrease both the depth $L$ and the backward connectivity $\xi $
is evident. In the most of evolutions where the average frequency $\langle
\omega \rangle $ has increased, a decrease in $L$ and/or $\xi $ was
observed. In Fig.~\ref{fig:learn}B, evolution of two other topological
properties is shown. We see that the correlation between the feedforward
connectivity $\chi $ and the mean frequency $\langle \omega \rangle $ is
weak. An increase in $n_{\mathrm{out}}$ indicates the emergence of a hub
(i.e., of a node with a large number of outgoing connections), which
strongly contributes to decreasing the depth $L$. It is worth noticing that
an increase in $\langle \omega \rangle $ does not necessarily imply an
increase in $n_{\mathrm{out}}$. Apparently, the entrainment ability relies
on more fine properties in the network architecture, best characterized
through $L$ and $\xi $. In our numerical simulations, we have tried several
sets of parameter values and several different initial random networks,
always obtaining qualitatively the same results. 

\emph{The second evolution algorithm}. For each iteration step, an oscillator 
$i$ is randomly chosen. A numerical simulation is run and the long-time
frequency of this oscillator, $\omega _{i}=\left[ \phi _{i}(2T)-\phi _{i}(T)%
\right] /T$ with sufficiently large $T$, is determined. Then, a structural
mutation of the network is introduced. We choose randomly one existing
incoming link of this oscillator ($A_{ij}=1$), delete it ($A_{ij}\rightarrow
0$), then choose randomly a missing link to this oscillator ($A_{ij^{\prime
}}=0$ where $j^{\prime }\neq i$) and add a new link there ($A_{ij^{\prime
}}\rightarrow 1$). After the mutation, a numerical simulation is repeated
and the new frequency $\omega _{i}^{\prime }$ of the oscillator $i$ is
measured. The mutation is accepted if $\omega _{i}^{\prime }>\omega _{i}$,
and rejected otherwise. At the next step, we again randomly choose an
oscillator and repeat the same procedure. Thus, in this evolutionary
process, the mutation is done according to the individual activities of the
oscillators. Note also that, in this evolution algorithm, the total number
of incoming connections for each oscillator is maintained constant.

\begin{figure}[tbp]
\resizebox{8cm}{!}{\includegraphics{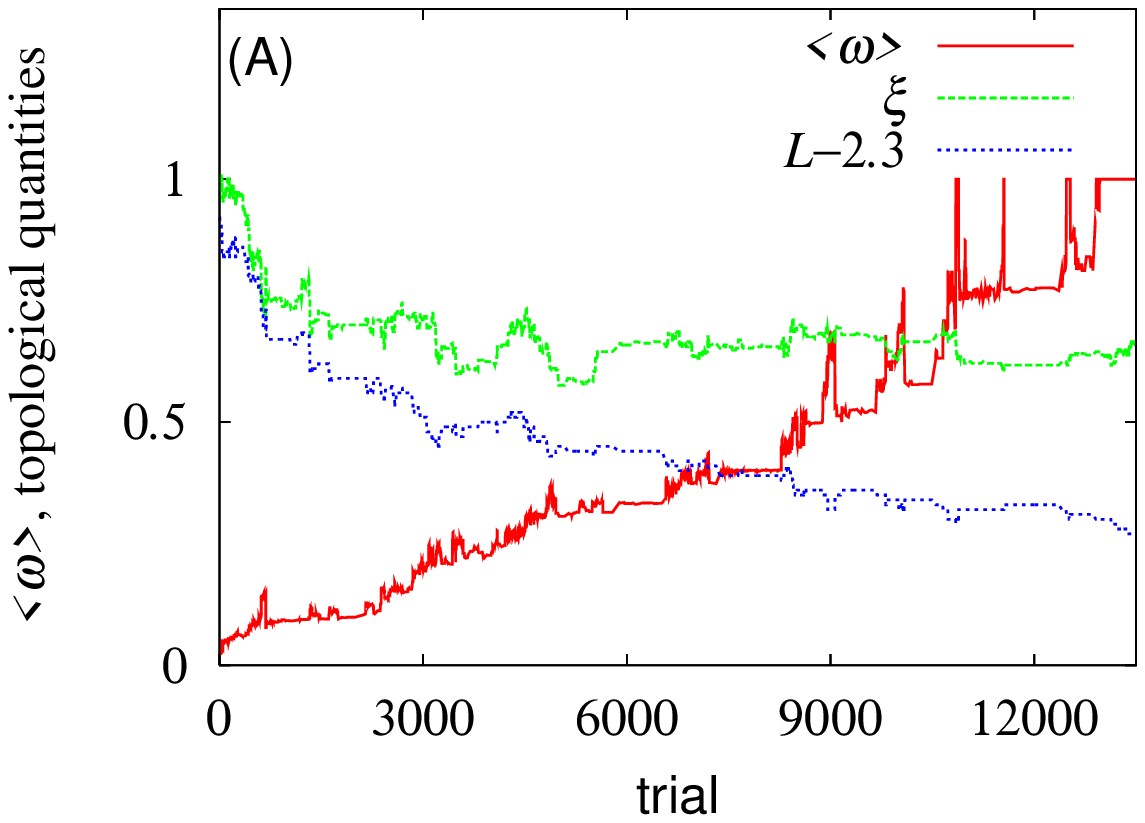}} %
\resizebox{8cm}{!}{\includegraphics{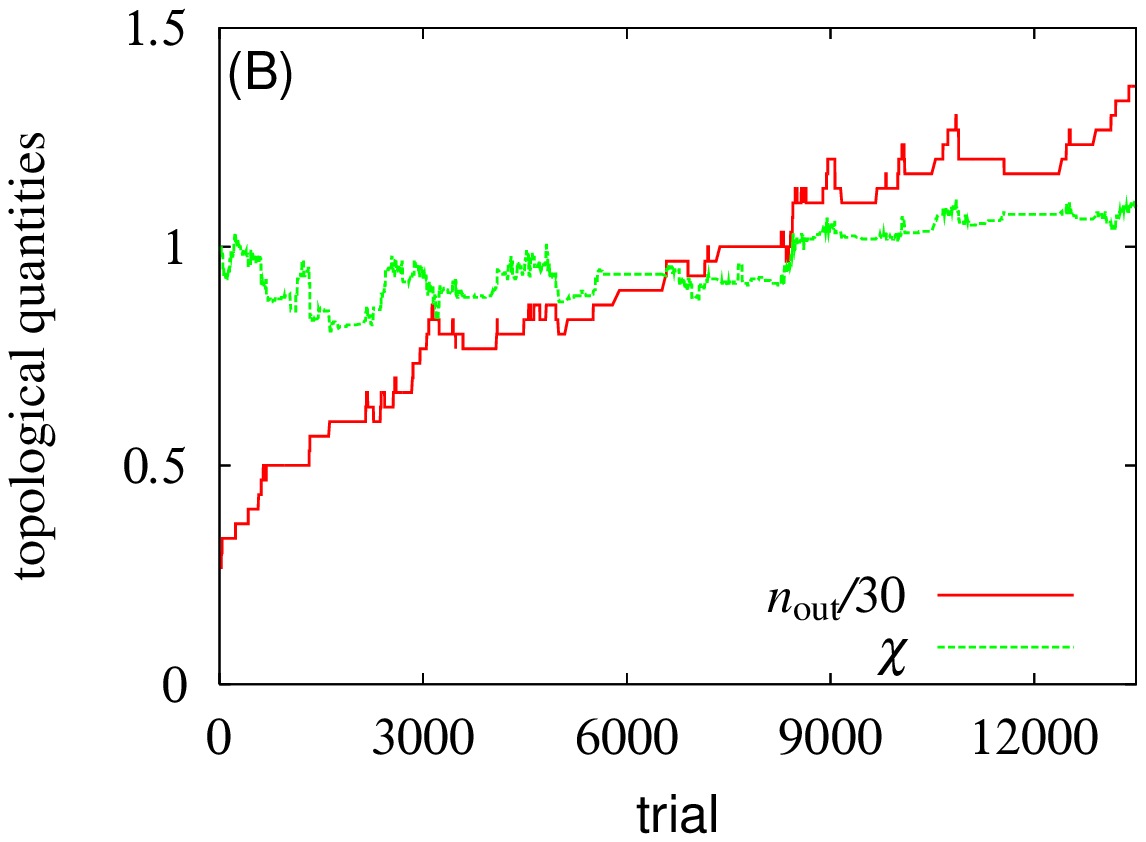}}
\caption{(color online). Evolution process under the second algorithm. The
depth of the initial random network is $L=3.22$, corresponding to $%
\kappa _{\mathrm{cr}}\simeq 100$. Parameter values and the definitions of
the exhibited lines are the same as in Fig.~\ref{fig:learn}.}
\label{fig:learn2}
\end{figure}

Figure~\ref{fig:learn2}A displays typical dependence of the average
frequency and two topological properties of the networks during a single
evolution. Although the average frequency $\langle \omega \rangle $ of
the whole population does not always increase in each iteration step,
the network architecture changes during the evolution similar to what
has been found for the first algorithm, towards the networks of smaller
depth $L$ and smaller backward connectivity $\xi $. In
Fig.~\ref{fig:learn2}B, emergence of a hub and weak correlation between
$\langle \omega \rangle $ and $\chi$ are seen, similarly to the results
in the first algorithm. It should be emphasized that, in the second
algorithm, globally coordinated network architecture emerges solely
through the \textquotedblleft judgments\textquotedblright of individual
oscillators based only on their own activity. The reason why this
\emph{local} optimization rule leads to the \emph{globally} coordinated
structure is that the oscillators are mutually frequency-synchronized in
most cases and therefore their individual frequencies usually coincide
with the average frequency of the whole population. Hence, the second
algorithm also works similarly to the first one.

In both types of algorithms, our numerical results show that the development
of the entrainment ability is followed by decreases in the depth $L$ and the
backward connectivity $\xi $, suggesting that these two topological
properties are the primary factors in determining the entrainment ability.
It can be also noticed that the evolving networks tend to develop strong
heterogeneity in their outgoing degrees. The depth of an evolving network
becomes smaller via an increase in the outgoing degrees of nodes in the
shallow shells. In this manner, hubs of outgoing connections are formed. On
the other hand, the development of small backward connectivity is mainly due
to a decrease of the outgoing degrees in the deep shells. Consequently, the
heterogeneity of outgoing degrees tends to be larger as the optimization
process goes on\footnote{%
Although there is only little similarity in the models, the evolutionary
process described in Ref.~\cite{ito02} has shown a similar result.}. Many
real-world networks, including scale-free networks, are known to have strong
heterogeneity in their degrees \cite{albert02}. Our results suggest that
evolving networks with good entrainment ability also become strongly
heterogeneous.

\section{Discussion of biological aspects}  \label{sec:discussion}

As already mentioned in Sec.~\ref{sec:introduction}, our study is motivated
by the SCN, the tissue orchestrating the circadian rhythms of the whole body
in mammals. In this section, we discuss the neural network structure of the
SCN and its possible roles from the viewpoint of the results of the present
investigations.

The SCN is anatomically organized into two groups, often called the "\textrm{%
core"} subdivision and the "\textrm{shell"} subdivision (see, e.g., Ref.~%
\cite{aton-review05}). Only the "core" subdivision receives photic input
and, thus, this part corresponds to the first shell in our model. The
"shell" subdivision corresponds to the rest of the shells in our model.
Within and between the "core" and the "shell" subdivisions, there are
several types of inter-cellular communications which influence circadian
oscillations \cite{aton-review05}. The pathways (i.e., the network
structure) of each type of communication may be different.

From the viewpoint of our results, if the SCN is constructed so as to
reach the best ability for entrainment, the network architecture should
be feedforward. This means that the unidirectional connectivity from the
"core" subdivision to the "shell" subdivision should be found. This is
indeed the case for communication via VIP (\emph{vasoactive intestinal
polypeptide}), which is a neurotransmitter released only by neurons in
the core subdivision. VIP is one of the leading candidate factors for
the synchronization pathway inside the SCN \cite{aton05}. However, VIP
is not the only one communication agent. Communication via GABA ($\gamma
$ -aminobutyric acid) is further possible for almost all neurons in the
SCN and seems to play a crucial role in achieving synchronization
between the "core" and the "shell" subdivisions \cite{albus05}. The GABA
connection pattern is not feedforward but bidirectional between these
subdivisions.  However, remarkably, it has been conjectured from
experimental studies that coupling from the "shell" to the "core" is
weaker than in the other direction \cite{albus05}, suggesting that the
network would effectively be close to feedforward one also with respect
to the GABA communication.

It is known moreover that, although the response of the "core" structure to
a sudden phase shift in the environment is very fast, the response of the
"shell" subdivision to such phase shifts is significantly slower \cite%
{nagano03,albus05}. From the viewpoint of our results, it can be conjectured
that the slow response of the "shell" subdivision is due to the more
hierarchical, non-feedforward organization of neurons inside this
subdivision. As we have seen, if a network is hierarchical, the response can
become very slow even if the first shell is strongly connected to the
environment.

It is interesting to ask why the "shell" subdivision, coupled to the
"core", actually exists in the SCN and why this subdivision could be
more hierarchically organized. One reason may be that such an
organization is needed to keep the autonomy of the biological clock. The
SCN must be capable to synchronize to the environmental rhythm, but, on
the other hand, should not be too sensitive to the environmental
information. In other words, the SCN must possess both autonomy and
adaptivity in a good balance. The hierarchical organization and
directionality of the network architecture in the SCN could be useful to
reconcile these contradicting requirements.

\section{Conclusions}

\label{sec:conclusion} 
\begin{figure}[tbp]
\resizebox{6.5cm}{!}{\includegraphics{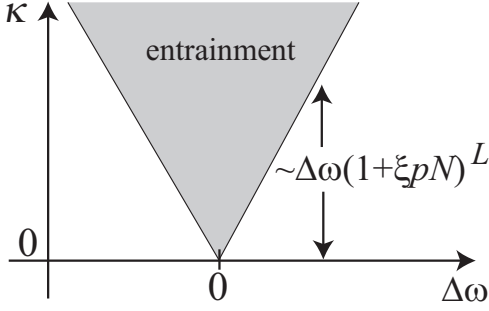}}
\caption{Schematic representation of the entrainment window, inside which
the entrainment occurs. The slopes of the edges of the entrainment window
for $\omega >0$ and $\omega <0$ are proportional to $\max {%
\Gamma }$ and $\min {\Gamma }$ respectively.}
\label{fig:tongue}
\end{figure}

Our main result for the random ER networks and
the random directionally biased networks is schematically illustrated in
Fig.~ \ref{fig:tongue}, where $\kappa $ is non-rescaled coupling
strength, i.e., $ \kappa $ in the original model (\ref{model}). The
entrainment occurs in a gray region. For the ER networks, this region
becomes exponentially smaller for the networks of higher hierarchical
organizations (i.e., with larger depth $L$). Thus, in practice, the
entrainment is possible only for shallow networks. For the directionally
biased networks, the entrainment region becomes significantly enlarged
for the smaller backward connectivity $\xi$ even if the networks are
hierarchical. The feedforward network is the best one for the
entrainment. The relaxation time to the entrainment has approximately
the same dependence on the network architecture as that of the
entrainment threshold. These results are general and hold for a large
class of coupled phase-oscillator systems (and thus also of weakly coupled
limit-cycle oscillator systems) with attracting couplings. The networks
with such topological properties are shown to emerge naturally through
different kinds of evolutions aimed at increasing the entrainment
ability.

The authors acknowledge fruitful discussions with K.~Honma, S.~Honma,
Y.~Kuramoto, N.~Masuda, F.~Mori., H.~Nakao, H.~Okamura, and
D.~Zanette. H.K. acknowledges financial support from the Humboldt
Foundation (Germany) and from the Center-of-Excellence Program
\textquotedblleft Nonlinear Structure via Singularity\textquotedblright\
in the Department of Mathematics of the Hokkaido University (Sapporo,
Japan).

\appendix

\section{Analytic investigations without the global tree approximation}

\label{sec:withouttree} In Sec.~\ref{sec:analytical}, to solve the
entrainment problems we have used the heuristic global tree approximation
for the random ER networks. By a different method, entrainment problems can
be solved directly for the ER network of large size $N\gg 1$ and large
degree $pN\gg 1$. Although this systematic derivation is more accurate, it
is also more technical and, therefore, we have preferred to present first,
in the main part of the article, the analysis based on the global tree
approximation. The results of this derivation agree with what we have found
before in the global tree approximation and can be viewed as providing
further support for it.

We begin by estimating the number $m_{hk}$ of incoming connections leading
from all nodes in the $k$th shell to a node in the $h$th shell. The network
pattern of forward connections is first considered, i.e., $m_{h,h-1}$ for $%
h\geq 1$ is estimated. By definition, there are $N_{1}$ nodes in the first
shell, each of which receives only one forward connection (coming from the
pacemaker). We thus have $m_{1,0}=1$. The expected total number of outgoing
connections from the nodes in the first shell is $N_{1}pN$ because each node
gives typically $pN$ outgoing connections. If $pNN_{1}\ll N$, every outgoing
connection almost surely connects to each individually different node
outside the first shell. Therefore, in a good approximation, the number $%
N_{2}$ of nodes in the second shell is $N_{1}pN$, and each node in the
second shell receives only one forward connection, i.e., $m_{2,1}=1$. The
same property holds up to a certain shell $h^{\ast }$, where for the first
time the total number of outgoing connections from the shell becomes of the
order $O(N)$ or larger, i.e., $pNN_{h^{\ast }-1}=o(N)$ and $pNN_{h^{\ast
}}\geq O(N)$. We thus have $N_{h}=N_{1}(pN)^{h-1}(\ll N)$ and $m_{h,h-1}=1$
for $h\leq h^{\ast }$. In other words, the network pattern of forward
connections takes a tree structure in a good approximation up to the shell $%
h^{\ast }$. The population $N_{h^{\ast }+1}$ of the next shell is of the
order of $O(N)$. Since $\sum_{h\leq h^{\ast }}N_{h}$ is of $o(N)$, almost
all outgoing connections from the shell $h^{\ast }$ connect to nodes in the
shell $h^{\ast }+1$. Therefore, the expected number of incoming connections
from the shell $h^{\ast }$ to a node in the shell $h^{\ast }+1$ is $%
pNN_{h^{\ast }}/N_{h^{\ast }+1}\equiv \bar{m}_{h^{\ast }+1,h^{\ast }}$. The
statistical deviation from this expected number may not be neglected because 
$\bar{m}_{h^{\ast }+1,h^{\ast }}$ is not a large number in general. Since $%
N_{h^{\ast }+1}=O(N)$, every node in the network almost surely receives
connections coming from the shell $h^{\ast }+1$, which implies that the rest
of nodes belong to the shell $h^{\ast }+2$. The expected number $\bar{m}%
_{h^{\ast }+2,h^{\ast }+1}$ of forward connections to a node in the last
shell is $pN_{h^{\ast }+1}$. This is a large number of $O(pN)$, and thus the
statistical derivation from the number can be neglected, i.e., $m_{h^{\ast
}+2,h^{\ast }+1}=pN_{h^{\ast }+1}$.

Since $L=\sum_{h}hN_{h}/N\simeq \{(h^{\ast }+1)N_{h^{\ast }+1}+(h^{\ast
}+2)N_{h^{\ast }+2}\}/N\simeq h^{\ast }+1+N_{h^{\ast }+2}/N$, we get 
\begin{equation}
h^{\ast }=[L]-1,  \label{app:h*}
\end{equation}%
where $[L]$ is the integer part of $L$. According to the result in \cite%
{fronczak04}, the depth of the ER random network with large size $N$ is
estimated as 
\begin{equation}
L\simeq \frac{\ln (N/N_{1})-\gamma }{\ln (pN)}+1.5,  \label{app:depth}
\end{equation}%
where $\gamma \simeq 0.5772$ is the Euler constant.

Next, the structure of backward and intra-shell connections is considered. A
node in the shell $h$ typically receives $pN_{k}$ connections from nodes in
the shell $k$ ($k\geq h$). The numbers of such connections from the last two
shells are of $O(pN)$. Relative statistical deviations from these numbers
are of order $(pN)^{-1/2}$ and thus negligible. We thus obtain $%
m_{h,k}=pN_{k}$ for $k=h^{\ast }+1,h^{\ast }+2$ and $h\leq k$. The number of
backward connections from other shells is of $o(pN)$, and thus negligible,
i.e., we approximate $m_{h,k}=0$ for $h<k\leq h^{\ast }$.

Our estimation for the network structure is summarized as follows. The shell
populations $N_{h}$ are $N_{1}(pN)^{h-1}$ for $1\leq h\leq h^{\ast }$ and of
order $O(N)$ for $h=h^{\ast }+1$ and $h^{\ast }+2$. The number $m_{h,k}$ of
incoming connections of a node in the shell $h$ from the nodes in the shell $%
k$ is given by $m_{h,h-1}=1$ for $1\leq h\leq h^{\ast }$ and by $\bar{m}%
_{h^{\ast }+1,h^{\ast }}=pNN_{h^{\ast }}/N_{h^{\ast }+1}$; $m_{h^{\ast
}+2,h^{\ast }+1}=pN_{h^{\ast }+1}$; $m_{hk}=0$ for $k<h-1$ (by definition); $%
m_{h,k}=pN_{k}$ for $k=h^{\ast }+1,h^{\ast }+2$ and $h\leq k$; and $%
m_{h,k}=0 $ for $h<k\leq h^{\ast }$.

Now we solve the entrainment problem for the network under consideration. We
assume that the phase difference $\phi _{i}-\phi _{j}$ between any pair of
oscillators in the shells $h\geq 2$ is so small that we may linearize as $%
\Gamma (\phi _{i}-\phi _{j})=\Gamma ^{\prime }(0)(\phi _{i}-\phi _{j})$.
This assumption is justified later [it will be shown that $|\phi _{i}-\phi
_{j}|<O(1/pN)$].

Dynamics in the shells $h>1$ is considered. We denote the average phase of
the oscillators inside the last two shells $h^*+1$ and $h^*+2$ by $\theta_{%
\mathrm{edge}}$. Because the number of connections from the last two shells
is large, we may approximate the coupling from those shells as 
\begin{equation}
\frac{1}{pN}\sum_j A_{ij} \Gamma^{\prime}(0)(\phi_i-\phi_j) \simeq
\Gamma^{\prime}(0)(\phi_i -\theta_{\mathrm{edge}}),  \label{appro-mean}
\end{equation}
where $j$ denotes the oscillators inside the last two shells. Since every
oscillator in the shell $h \le h^*$ has effectively the same number of
incoming connections from each individual shell, a state with phase
synchronization inside each shell $h \le h^*$ exists. We denote the phase of
the oscillators in the shell $h$ by $\theta_{h}$. Under entrainment (i.e., $%
\dot \phi_i=1$), the phases of such a state can be found as a solution of
algebraic equations 
\begin{equation}
\frac{\kappa}{pN} \Gamma^{\prime}(0) (\theta _{h}-\theta_{h-1}) + \kappa
\Gamma^{\prime}(0) (\theta_{h}-\theta_{\mathrm{edge}}) = 1,  \label{non-edge}
\end{equation}
where $2 \le h \le h^*$. The phase $\phi^{\mathrm{edge}}_i$ of the
oscillator $i$ inside the last two shells $h^*+1$ and $h^*+2$ are found by 
\begin{equation}
\frac{\kappa}{pN} \sum_j A_{ij} \Gamma^{\prime}(0) (\phi^{\mathrm{edge}%
}_{i}-\phi_j^{h^*}) + \kappa \Gamma^{\prime}(0) (\phi^{\mathrm{edge}%
}_i-\theta_{\mathrm{edge}}) = 1.  \label{edge}
\end{equation}
Averaging Eq.~(\ref{edge}) for all the oscillators inside the last two
shells, we get 
\begin{equation}
\frac{\kappa N_{h^*}}{N} \Gamma^{\prime}(0)(\theta_{\mathrm{edge}}-\theta_{%
\mathrm{h^*}})=1.  \label{edge2}
\end{equation}
Note that because of the strong internal coupling inside the last two
shells, the oscillators inside the last two shells are almost
phase-synchronized, i.e, $\phi^h_i \simeq \theta_{\mathrm{edge}}$ for $%
h=h^*+1$ and $h^*+2$. From Eq.~(\ref{non-edge}) and Eq.~(\ref{edge}), for $2
\le h \le h^*$ we obtain 
\begin{eqnarray}
\theta_{\mathrm{edge}}- \theta_{h^*}&=& \frac{N}{\Gamma^{\prime}(0) \kappa
N_{h^*}},  \notag \\
\theta_h - \theta_{h-1} &=& \frac{pN}{\Gamma^{\prime}(0) \kappa}
(1+pN)^{h^*-h} \left(1+\frac{N}{N_{h^*}} \right),  \notag \\
\Gamma(\theta_{2} - \theta_1) &=& \frac{pN}{\kappa} (1+pN)^{h^*-2} \left(1+%
\frac{N}{N_{h^*}} \right).  \label{sol}
\end{eqnarray}

The existence condition of the entrained solution are $\tilde
\Gamma(\theta_{1} - \theta_{0}) \le 1$ and $\Gamma(\theta_{2} - \theta_{1})
\le 1$. The former is satisfied for sufficiently large $\mu$ (which is our
assumption). Thus, for a given network, the existence condition for the
entrained solution is $\kappa \le \kappa_{\mathrm{cr}}$, where 
\begin{equation}
\kappa_{\mathrm{cr}} = \alpha^{-1} pN (1+pN)^{h^*-2} \left(1+\frac{N}{N_{h^*}%
} \right).  \label{app:kcr}
\end{equation}
Substituting $\kappa_{\mathrm{cr}}$ into $\kappa$ in Eqs.~(\ref{sol}), one
can confirm that the phase difference $\theta_2-\theta_{\mathrm{edge}}$ is
actually at most of $O(1/pN)$, so that the linear approximation is justified
for large $pN$.

In Eq.~(\ref{app:kcr}), because $h^{\ast }$ increases with the depth $L$,
the entrainment threshold approximately increases exponentially with the
depth $L$, although the effect of the last term $1+N/N_{h^{\ast }}$ is
unclear. We thus employ further approximations and express the entrainment
threshold (\ref{app:kcr}) as a function of the depth $L$ and the mean degree 
$pN$. From Eq.~(\ref{app:depth}), the network size can be expressed as $%
N=e^{\gamma }N_{1}(pN)^{L-1.5}$. Together with the relations $N_{h^{\ast
}}=N_{1}(pN)^{h^{\ast }-1}$ and $N\gg N_{h^{\ast }}$, the last term of Eq.~(%
\ref{app:kcr}) is approximated as 
\begin{equation}
1+N/N_{h^{\ast }}\simeq N/N_{h^{\ast }}\simeq e^{\gamma }(pN)^{L-h^{\ast
}-0.5}.
\end{equation}%
Because $0.5\leq L-h^{\ast }-0.5<1$ from Eq.~(\ref{app:h*}), we may further
approximate it as 
\begin{equation}
e^{\gamma }(pN)^{L-h^{\ast }-0.5}\simeq e^{\gamma }(1+pN)^{L-h^{\ast
}-0.5}+C,  \label{app:appro1}
\end{equation}%
where $C$ is a small number of order $(pN)^{L-h^{\ast }-1.5}$. Substituting
Eq.~(\ref{app:appro1}) into Eq.~(\ref{app:kcr}) and neglecting the second
term $C$ which is much smaller than the first term, we get 
\begin{equation}
\kappa _{\mathrm{cr}}\simeq e^{\gamma }pN(1+pN)^{L-2.5}.  \label{app:kcr2}
\end{equation}%
This entrainment threshold agrees well with numerical data without any
fitting parameter. Dividing Eq.~(\ref{app:kcr2}) by Eq.~(\ref{kcr}), we
obtain $e^{\gamma }(1+pN)^{-0.5}\equiv c$, which is about $0.67,0.54$ and $%
0.39$ respectively for $pN=6,10$ and $20$. These agree well to $c=0.67,0.60$
and $0.45$ obtained by numerical fitting (see Fig.~\ref{fig:kcr-other}).

\section{Entrainment threshold for bidirectional networks}

\label{sec:bidirectional} So far, we have considered only directed networks.
However, similar exponential dependences are found also for bidirectional
networks. In this section, we provide estimation of the entrainment
threshold for \emph{bidirectional} ER networks. Such networks are generated
as follows. Independently for any $i$ and $j<i$, we set $A_{ij}=1$ with
probability $p$ and $A_{ij}=0$ otherwise, and put $A_{ji}=A_{ij}$.
Self-connections are forbidden, so that $A_{ii}=0$.

The entrainment threshold is obtained in a similar way as in Sec.~\ref
{sec:analytical}. We apply a global tree approximation in forward
connections (which implies that the pattern of backward connections also
take the same structure). Every oscillator then receives only one
forward connection. By assuming $pN\gg 1$, every oscillator (except
those in the last shell) receives approximately $pN$ backward
connections \footnote{ This approximated network is the same as the
Cayley tree. Using some particular phase models, the entrainment
solution for the Cayley tree network has been obtained by Yamada
\cite{yamada02} and Radicchi and Meyer-Ortmanns
\cite{radicchi06}.}. Therefore, the only difference from the ER directed
networks is that backward connections come not from the last shell but
from the next shell. Thus, we get the following equations:
\begin{equation}
\mu \tilde{\Gamma}(\theta _{1}-t)+\kappa \Gamma (\theta _{1}-\theta _{2})=1,
\label{reducedmodel-bi-1}
\end{equation}%
\begin{equation}
\frac{\kappa }{pN}\Gamma (\theta _{h}-\theta _{h-1})+\kappa \Gamma (\theta
_{h}-\theta _{h+1})=1\quad \mbox{for $2\le h <L $},
\label{reducedmodel-bi-2}
\end{equation}%
\begin{equation}
\frac{\kappa }{pN}\Gamma (\theta _{L}-\theta _{L-1})=1.
\label{reducedmodel-bi-3}
\end{equation}%
Using the approximations available for large $pN$, we obtain 
\begin{equation}
\Gamma (\theta _{h}-\theta _{h-1})=\frac{pN\{(pN)^{L-h+1}-1\}}{\kappa (pN-1)}%
\quad \mbox{for $h \ge 2 $}.  \label{phase-bi}
\end{equation}%
The entrainment thresholds is thus 
\begin{equation}
\kappa _{\mathrm{cr}}=\frac{pN\{(pN)^{L-1}-1\}}{pN-1}\simeq (pN)^{L-1}.
\label{kcr-bidirectional}
\end{equation}%
Stability analysis can also be performed in a same manner as in Sec.~\ref%
{sec:stability}. We then get the same relaxation time as Eq.~(\ref{tau}). It
is thus found that the dependences of the entrainment threshold and the
relaxation time are also exponential in the bidirectional networks, but
their functional forms are slightly different from those for the directed
networks.

\end{document}